\documentclass[11pt]{article}

\usepackage{graphicx}
\usepackage{amssymb}
\usepackage{amsmath}
\let\UnmodifSec=\section
\renewcommand{\section}{\setcounter{equation}{0}\UnmodifSec}

\def\dateline{}
\usepackage{esvect}
\usepackage{xcolor}

\def\kk{\kappa}
\def\x{{\rm x}}
\def\k{{\rm k}}
\def\z{{\rm z}}
\def\t{\y^0}

\def\l{\lambda}
\newcommand{\kkappa}{R}

\newcommand{\iu}{\ensuremath{i}}

\newcommand{\AdS}{\ensuremath{AdS_d}}

\newcommand{\AdSC}{\ensuremath{AdS_d^{(c)}}}
\newcommand{\scalar}[2]{\ensuremath{#1\cdot #2}}

\newcommand{\zz}{{\mathbf{z}}}

\newcommand{\zu}{z}

\def\K{\varkappa}

\def\W{\mathfrak{W}}

\renewcommand{\cosh}{\mbox{ch\,}}\renewcommand{\sinh}{\mbox{sh\,}}
\renewcommand{\tanh}{\mbox{th\,}}\renewcommand{\tan}{\mbox{tg\,}}
\def\dil{{D}}

\def\bC{{\bf C}}
\def\bR{{\bf R}}

\def\bZ{{\bf Z}}
\def\Im{{\rm Im\,}}
\def\Re{{\rm Re\,}}
\def\Rp{{\bf R}_+}

\def\CC{{\cal C}}

\def\GG{{\cal G}}
\def\HH{{\cal H}}

\def\MM{{\cal M}}

\def\TT{{\cal T}}

\def\WW{{\cal W}}

\def\ZZ{{\cal Z}}
\def\wh{\widehat}
\def\wt{\widetilde}
\def\ovl{\overline}

\def\interior#1{\setbox1=\hbox{$#1$}\rlap{$#1$}\kern0.4\wd1\raise1.1\ht1%
\hbox{$\scriptstyle \circ$}}

\def\boxit#1#2{\setbox1=\hbox{\kern#1{#2}\kern#1}%
\dimen1=\ht1 \advance \dimen1 by #1 \dimen2=\dp1 \advance \dimen2 by #1
\setbox1=\hbox{\vrule height\dimen1 depth\dimen2\box1\vrule}%
\setbox1=\vbox{\hrule\box1\hrule}%
\advance \dimen1 by .4pt \ht1=\dimen1 \advance \dimen2 by .4pt \dp1=\dimen2
\box1\relax}
\def\endprf{\raise .5ex\hbox{\boxit{2pt}{\ }}}
\def\p{{\rm p}}
\def\zz{{\rm z}}
\def\amb{{\bf R}^{d+1}_{2}}
\def\ambc{{\bf C}^{d+1}_{2}}

\def\ifundefined#1{\expandafter\ifx\csname#1\endcsname\relax}

\title{\bf Two point functions and quantum fields in the anti-de Sitter universe\footnote{Dedicated to the memory of Henri Epstein}}
\author {Ugo Moschella\\[20pt]
{\small Dipartimento di Scienza e Alta Tecnologia, Via Valleggio 11, 22100 Como,} \\
{\small  INFN sez. di Milano, Via Celoria 6, 20133 Milano, Italy}}

\date{\dateline}

\begin{document}

\maketitle

\begin{abstract}
We construct a manifestly covariant and coordinate-free plane-wave representation of scalar two-point functions in 
$d$-dimensional anti-de Sitter spacetime. The construction is based on a new class of holomorphic plane waves defined globally on the universal covering of the AdS via chiral cones in the complex null cone. Imposing AdS invariance, locality, positive definiteness and a spectral condition, we obtain integral representations as superpositions over relative homology cycles, reproducing the standard maximally analytic solutions in terms of Legendre functions of the second kind. In Poincaré coordinates, the two-point functions diagonalize into a K\"all\'en-Lehmann superposition of 
$(d-1)$-dimensional Minkowski correlators where the weight is a product of Bessel functions. This diagonalization clarifies the relation between Euclidean and Lorentzian AdS quantum field theory and allows Wick rotation of Euclidean Feynman diagrams to Lorentzian integrals supported on a single Poincaré patch while preserving full AdS covariance.
\end{abstract}

\section{Introduction}

The concept of a cyclic universe is deeply rooted in many cultures and long predates the notion of a linear, historical cosmos. In the Western tradition, Pythagoras, who is credited with introducing the term {“cosmos"}, was also the first  to speak of a universe governed by recurring cycles. Porphyry, in his {\em Life of Pythagoras}, conveys the Pythagorean cosmological beliefs explicitly referring to the idea of cosmic cycles \cite{porph}:
{\em “He taught that [...] after certain specified periods, the same events occur again.”}  In Eastern traditions, the idea of the universe as a cycle is often even more deeply ingrained. Hindu cosmology speaks of the Great Era, a vast cycle of creation, preservation, and destruction, while Buddhist thought highlights endless ages of arising and dissolution. Similarly, Chinese philosophy envisions the cosmos as continuously renewing itself through alternating cycles.
Mircea Eliade, in his book {\em Le mythe de l'\'eternel retour}  observed that virtually all ancient societies conceived of time as an  endlessly recurring cycle, and rejected the notion of a singular, linear historical timeline \cite{eliade}: {\em “No event is unique, occurs once and for all... but it has occurred, occurs, and will occur perpetually; the same situations are reproduced that have already been produced in previous cycles and will be reproduced in subsequent cycles — ad infinitum."}

 From a branch of philosophy and religion, cosmology evolved into a scientific discipline very recently, with the advent of Einstein's general theory of relativity in 1915. In his initial cosmological model of 1917, Einstein assumed the universe to be static and unchanging, much like the skies  of Ptolemy and Copernicus. To eliminate the notion of absolute space, he represented the universe as spherical, which led him to incorporate the cosmological constant—the key term to ensure  staticity of a spherical  dusty universe.  Edwin Hubble's 1929 observations later showed that the universe is actually expanding and consistent with Einstein's original equations and therefore {\em if there is no static universe  away with the cosmological constant} \cite{strautmann}.  Then, in 1998, two teams, independently studying distant  supernovae, found that the expansion of the universe is  accelerating. This surprising discovery  revived interest in the cosmological constant and suggested that it may constitute the dominant component of the universe's energy content.

If the cosmological constant is truly nonzero, could it be negative? A negative cosmological constant entails fundamentally different physics: the vacuum energy density is negative and the gravitational effect attractive, in contrast to the repulsion caused by a positive cosmological constant. Recent research \cite{neg,Menci} however indicates that a negative cosmological constant, when combined with other dark energy components, can be compatible with observational data, potentially allowing for a broader range of cosmic evolutions.

A universe devoid of matter and energy except for a negative cosmological constant has the anti-de Sitter (AdS) geometry. It features a constant negative curvature and exhibits closed timelike curves; while the light-cone ordering holds locally, it breaks down globally. In this respect, the AdS universe has a more evocative resemblance to the old cyclical cosmological worlds of Pythagoras  rather than the modern relativistic expanding universes of  Friedmann and Lemaître. 

While the  cyclical vision of time provided ancient peoples with a sense of continuity and meaning beyond the relentless progression of historical events, it is perceived as a scandal by  physicists   who are deeply committed to the causality principle and find such cyclicality at odds with the very idea of physical law:
 {\em “many physicists are unhappy with the closed timelike curves  and seek to assuage their feelings of guilt by claiming to pass to the universal covering spacetime. In this way they feel that they have exorcised the demon of “acausality”. However, therapeutic uttering these words may be, nothing is actually
gained in this way."} (quoted from: G. Gibbons \cite{gibbons}).

\begin{figure}[ht]
  \centering
    \includegraphics[width=0.5
    \textwidth]{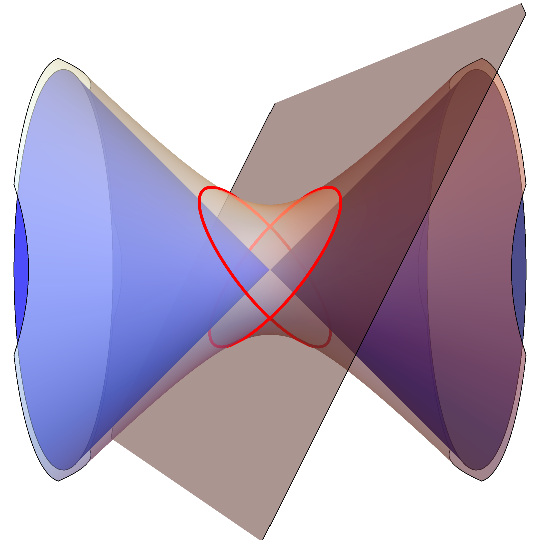}
  \caption{Timelike geodesics are periodic curves. The planes containing these geodesics do not intersect the real null cone. Each timelike geodesic may be parametrized by a single complex null vector. All timelike geodesics originating from a given point reconverge at its antipodal point after half a period. This focusing phenomenon persists in the universal covering manifold as well. }
  \label{ads.geo2}
\end{figure}

 The presence of closed timelike curves in AdS spacetime constitutes a genuine conceptual challenge in physical modeling.
 Calling it a scandal is perhaps an overstatement but, nevertheless, the word highlights a real and significant tension with causality.
The challenge remains unresolved by passing to the universal covering spacetime:   the geodesics originating from one point  converge  at the antipodal point after a half-period of {\em “six great months"} and again at the starting point after a full {\em "great era." } The universal cover removes the identification of points along the closed timelike curves but the periodic focusing of geodesics persists, providing no real gain in resolving the  issues related to cyclic time in AdS (see Figure \ref{ads.geo2}).

This conceptual challenge explains why even apparently simple questions—such as finding an explicitly AdS-covariant expression for the correlation functions of free quantum fields—have remained unresolved to date. In the standard approach to AdS quantum field theory, which mainly relies on imposing boundary conditions at spacelike infinity for modes decomposed in specific coordinate systems \cite{Fronsdal,avis,breit}, the problem seems intractable. This method tends to obscure the global analyticity properties of both the AdS manifold and the correlation functions \cite{bem}. Neglecting these properties implies that the difficulties introduced by closed timelike curves could only be overcome with substantially greater effort.

This is one of the issues we tackle in the present work. We develop a new representation for the two-point functions of scalar quantum fields in a $d$-dimensional anti-de Sitter universe, based on a novel family of holomorphic plane waves that we introduce here and previously announced in  \cite{mosk}.  These plane waves are almost always quasi-periodic functions on the AdS manifold and naturally reside on its covering space. The sole exception arises when the mass parameter takes integer values; in such cases, the plane waves become genuinely periodic.

Here is,  deferring all detailed explanations, our formula for the two-point function of a massive scalar field: \begin{eqnarray}  W^{(d)}_\lambda(z_1,z_2)&=&  c_d(\lambda) \int_{\gamma(z_1)} (z_1\cdot \zeta)^\lambda ( z_2\cdot \zeta)^{-\lambda-d+1} d\mu_\gamma(\zeta)= \label{pwduint}\\
&=&  c_d(\nu) \int_{\gamma(z_1)} (z_1\cdot \zeta)^{-\frac{d-1}{2}+\nu} ( z_2\cdot \zeta)^{-\frac{d-1}{2}-\nu} d\mu_\gamma(\zeta). \label{pwduintnu}
\end{eqnarray}
This integral representation is a superposition of plane waves; it does not separate space and time variables, nor does it depend on any particular coordinate system on the AdS manifold or any specific class of special functions. It is explicitly invariant under the anti-de Sitter group and clearly reveals the primitive analyticity domain of the two-point function \cite{bem}. Correlation functions for tensors and spinors can be straightforwardly derived by applying suitable differential operators to the right-hand side of this expression.

The significance of the new representation (\ref{pwduint}) is evident: it provides the closest possible analogue, within the AdS framework, to the plane-wave Fourier representation of the two-point function for a massive scalar field in (complex) Minkowski space $M_d^{(c)}$ with positive energy spectrum \cite{pct} which we quote here 
\begin{eqnarray}\label{tp}
 {W}^{{ M}_d}_{m}({\zz}_1,{\zz}_2) =
\frac{1}{(2\pi)^{d}}\int  e^{-i \p\cdot \zz_1} e^{i \p\cdot \zz_2}\theta(\p^0)\delta(\p^2-m^2)d\p. \label{tpm}
\end{eqnarray}
and to the plane wave expansion for the de Sitter quantum fields we have constructed  long ago \cite{bgm,bm}. Manifestly covariant momentum space representations of quantum fields provide  the physically intuitive ingredients of Feynman diagrammatics \cite{veltman} and a great deal of the predictive power of  quantum field theory depends on them. Furthermore, many crucial mathematical developments such as the  PCT and the Spin-Statistics theorems \cite{pct} and the Euclidean formulation of quantum field theory \cite{euclidean} are based on the plane-wave Fourier representation of  the correlation functions and the positivity of the energy spectrum. The Fourier representation (\ref{tp}) and the related  representation of the Feynman propagator are of paramount importance. We expect (\ref{pwduint}) to be just as fundamental in the framework of AdS QFT.

The second major topic addressed in this work pertains to the horospherical, or Poincar\'e, coordinate system, which covers exactly one half of the AdS manifold. We analyze the Fourier transform of the plane waves w.r.t. the Poincar\'e coordinates and show how  this transform diagonalizes the two-point function (\ref{pwduint}). We present here the final result, which exhibits remarkable elegance, while deferring all detailed derivations to Sect.~\ref{sec13} and \ref{sec14}:
\begin{eqnarray}
&&W^{(d)}_\lambda(z_1(\z_1,u),z_2(\z_2,u')= \cr &&=
\frac{\left(u u'\right)^{\frac{d-1}{2}}}{2 (2\pi) ^{d-2} }
   \int e^{-i \k(\z_1-\z_2)} \theta(\k^0) \theta(\k^2) J_{\frac{d-1}{2}+\lambda }\left(u \sqrt{\k^2} \right) J_{\frac{d-1}{2}+\lambda }\left(  u'\sqrt{\k^2}\right)d\k. \label{poinc}
\end{eqnarray}

{Some historical remarks on this integral representation may help to put it in perspective. The Euclidean counterpart of Eq.~(\ref{poinc}), which represents the Schwinger propagator, appears in Bateman's tables of integral transforms \cite[Eq.~(12), p.~64]{batemantransform2}. There, a Legendre function of the second kind \cite{Bateman}, with a rather intricate argument, is identified as the Hankel transform of a product of two Bessel functions.

A particular case of Eq.~(\ref{poinc}), corresponding to  spacetime dimension 
$d=2$, is discussed in Watson's classic treatise \cite{watson}, together with indications of the proof. A similar integral—Watson notes—had already been discovered by Kirchhoff as early as 1853. Watson also credits other particular cases to \cite{beltrami,Sommerfeld,Gegenbauer,macdonald} and refers to all of them as Lipschitz–Hankel integrals. Formula~(\ref{poinc}) may therefore be called a Lipschitz–Hankel formula \cite{maier}, or simply a Hankel formula, as we shall do throughout the rest of this paper.

In the AdS literature, the Euclidean Hankel formula for the Schwinger propagator was reported in an appendix \cite[Eq.~(A.3)]{tsi}, where it appears as an intermediate step toward reconstructing a more conventional representation of the Feynman propagator, always in terms of  Bessel functions, previously introduced by \cite{mueck}.

The geometrical interpretation in anti-de Sitter (AdS) space of the Hankel formula (\ref{poinc}), together with its relation to the maximally analytic Wightman functions of AdS quantum field theory \cite{bem}—which also include the Euclidean Schwinger propagator—was established shortly thereafter in \cite{bertola1,bertola2}. This result was obtained by employing complexified Poincaré coordinates and exploiting the general analyticity properties of Wightman functions that satisfy the AdS spectral condition \cite{bem}.
From a technical perspective, the Hankel formula was derived in \cite{bertola1} by solving a Schrödinger-type equation in the spacelike coordinate transverse to the Poincaré foliation. A key ingredient in this derivation is Bateman’s formula, whose geometrical interpretation was clarified in that work. This approach also provides insight into the so-called Breitenlohner–Freedman phenomenon \cite{breit}. In fact, the operator appearing in the transverse Schrödinger problem is not essentially self-adjoint, and its possible self-adjoint extensions correspond to different admissible quantizations in the regime where two such schemes coexist, thereby offering an alternative interpretation of the Breitenlohner–Freedman phenomenon.

The Lorentzian Hankel formula for the Feynman propagator, which had been somewhat overlooked for a period of time, was eventually revived and brought to general attention in \cite{raju,raju2}. These studies not only reintroduced the formula but also demonstrated its significant utility in calculations of propagators, correlation functions, and scattering amplitudes in the AdS background. Since then, the formula has become  part of the standard AdS toolbox.

Here, the Hankel formula and its untapped potential role in AdS QFT are brought into sharper focus. In particular, its connection to the complex structure of the AdS manifold, the geometry of the complex cone, and the plane-wave solutions of the Klein–Gordon operator is fully elucidated. This perspective highlights the crucial role played by the global analyticity properties of the plane waves, which are inherited from the AdS spectral condition \cite{bem}, and makes their significance evident in the context of the new derivation.

Moreover, our analysis uncovers an unexpected and intriguing insight. It demonstrates that Euclidean Feynman diagrams constructed in Lobachevsky space can, through a standard Wick rotation, be mapped onto diagrams entirely supported within the Poincar\'e patch of AdS. Several concrete examples are provided that support the conjecture that any Feynman or Witten diagram, computed by integration solely over a Poincaré patch of the real AdS manifold, continues to respect full AdS covariance, even though the Poincaré patch itself is not invariant under the full isometry group.}

\vskip 10 pt

The structure of the paper is as follows: in Section \ref{PRE}, we describe the geometric and group-theoretical foundations, detailing  the real AdS spacetime as an embedded manifold, its complexification, and the universal covering of the real manifold. 

In Section \ref{conepw} we introduce the null cone in the complex ambient space and explore its connection to geodesic motion in the AdS universe  following the approach developed in \cite{cacc} for the geometrically simpler de Sitter case.

In Section \ref{general} we describe the chiral tuboids originally introduced in \cite{bem}: these are complex analytic domains that play a pivotal role in the study of quantum fields on AdS spacetime. Within our framework, analyticity in these tuboids replaces the conventional approach, which relies on imposing boundary conditions at spacelike infinity on individual modes. 

Section \ref{axiom} investigates the analytic structure of two-point functions, which must meet the fundamental requirements of AdS covariance, locality, and positive definiteness—essential principles for any quantum field theory.
However these general properties alone do not uniquely determine the solutions to the field equations of a given model; additional physical input is necessary to distinguish physically meaningful solutions from the vast set of mathematical possibilities. Here, as in Minkowski space, this role is played by a spectral condition, providing both a physical and a mathematical criterion for quantization. For two-point functions the spectral condition is equivalent  to the maximal analyticity of the correlation functions in a certain complexification of the  AdS manifold.

In Section \ref{dal}, we construct plane wave solutions to the  AdS-Klein-Gordon equation  by exploiting the relationship between the pseudo-Laplacian in the ambient space and the AdS d'Alembertian. Plane waves are labeled by null vectors in the ambient space and by a mass parameter; in general they are well-defined only locally; the introduction of chiral cones in Section \ref{cone} provides a means to extend them into  objects globally well-defined on the universal covering of the AdS manifold.  Some mathematical aspects of our chiral cones are provided in Appendix \ref{Appa}.

In Section \ref{kgf}  we return to the global analytic structure  of the two-point functions for Klein-Gordon fields and recall their expression  in terms of Legendre functions of the second kind whose argument is the scalar product of two complex AdS events, lifted to a complex domain of the Cartesian product of two copies of the covering manifold. 

 Section  \ref{sec9} and \ref{sec10} focus on $AdS_2$
  as a testing ground for the general case.  We provide a detailed proof of the plane-wave expansion (\ref{pwduint}) and derive several novel integral representations of the two-point functions previously described in Section \ref{kgf}. The main challenge is the proper identification of the integration cycle $\gamma$ on the complex cone, a difficulty that reflects the intricate topological features of the AdS universe. Additionally, we address the bulk-to-boundary limit, which is crucial for holographic correspondences, and introduce the shadow transformation—a mathematical relation among dual waves of a given mass. These issues are resolved first in the two-dimensional context, which, while simpler, captures the critical aspects of the general case discussed in Section \ref{sec11}. Furthermore, we present in an example  the value of our new formula (\ref{pwduint}) to derive nontrivial theorems concerning Legendre functions of the second kind. 

In Section \ref{sec12}, we analyze the relevant integration cycles of the two-point function (\ref{pwduint}) within the particularly significant Poincaré  coordinate system. In the subsequent Sections \ref{sec13} and \ref{sec14}, we compute the horospherical-Fourier transform of the plane waves and show  how this transform can be utilized to diagonalize the two-point function. The calculation fully exploits the holomorphic structure of the plane waves and the connection between the ambient AdS chiral tuboids and the Minkowskian tubes of the Poincar\'e leaves.

This approach allows for clarifying the issue of Wick rotations illustrating the relation between Lorentzian and Euclidean calculation in AdS. In the last  Section \ref{sec15} we derive a  formula for the Feynman-Schwinger propagator corresponding to (\ref{poinc}) and show how all banana Euclidean Feynman diagrams can be Wick-rotated into banana diagrams in the real manifold, where the integration is restricted solely to a Poincare patch. We also explore examples of diagrams with external legs where the integration is again restricted  to a Poincar\'e patch and find that the result is AdS invariant.

The concluding remarks reflect on the broader significance of the results, their potential applications to perturbative quantum field theory in AdS, and the continuing evolution of both mathematical and physical understanding in this area.

Finally in Appendix A we discuss some technical details and in Appendix B we revisit the so called "split representation" of the Euclidean propagator

\section{AdS and its covering: geometrical setup}

\label{PRE}

Let $\amb$ denote\footnote{ The upper index in $\amb$ indicates the dimension of the real vector space and the lower index indicates the number of positive directions. The same convention applies to $\ambc$} the vector space $\bR^{d+1}$ equipped with the scalar product
\begin{eqnarray}
&&\scalar{x}{y}  = x^0 y^0 + x^d y^d - \vec{x}\cdot \vec{y} = g_{\mu\nu}x^\mu x^\nu\ ,
\label{2.1}\ \ \  g_{\mu\nu} = diag(+1,-1,\ldots, -1,+1);
\label{adsscp}\end{eqnarray}
the coordinates $x^\mu$ are relative to a chosen canonical  basis $\{e_0,e_1,\ldots, e_d\}$ for  $\amb$: 
\begin{equation}
      \ \ 
\scalar{e_\mu}{e_\nu}   = g_{\mu\nu},   \ \  x = x^\mu e_\mu, \ \   \vec{x} = (x^1,\ \dots,\ x^{d-1}). 
\end{equation} 
A vector 
$x$ %in $\amb$ 
is called timelike, spacelike or lightlike  (aka null) 
according to whether $\scalar{x}{x}$ is positive,
negative or null.

$G=O(2,d-1)$    is the orthogonal  group of all real linear transformations
of $\amb$ which preserve the scalar product (\ref{2.1}). $G$ has four components;  the component of $G$ connected to the identity is denoted by $G_0= SO^+(2,d-1)$ and is interpreted as the relativity group of the AdS universe, which can be identified with the manifold
\begin{equation}
\AdS=\{x,\ x^2=\scalar{x}{x}=R^2\}.
\end{equation}
The restriction to $\AdS$ of the pseudo-Riemannian metric
$g_{\mu\nu}dx^\mu dx^\nu$ is a Lorentzian metric with signature
$(+,\ -, \ldots, -)$. $G_0$  acts transitively on $\AdS$; we chose   \begin{equation}
    e_d = (0,\ldots,0,R)
    \end{equation}
    to be the distinguished origin of $\AdS$; in the following we will take  for simplicity a unit radius $R=1$.

The  complex $d-$dimensional anti-de Sitter  manifold $\AdSC$ can analogously be defined by its embedding in the complex  space  $\ambc$ endowed with the metric $g_{\mu\nu}$:
\begin{equation}
\AdSC=\{z=x+\iu y \in \ambc,\ z^2=z\cdot z=1\}
=\{z=x+\iu y,\ x^2-y^2=1,\ \scalar{x}{y}=0\}.
\end{equation}
The complexification $G_0^{(c)}$ of the real AdS group acts
transitively  $\AdSC$.

In spacetime dimension  $d \ge2$ the {covering space} of $\AdSC$ 
is $ \AdSC$ itself.  
However,  although the full space   
$ \AdSC$ itself has a trivial covering, $G_0$ and $AdS_d$ admits nontrivial 
{covering group }  $\wt G_0$ and {covering space } 
$\wt \AdS$; also, the important tubular domains  $\ZZ_\pm$ of   
$ \AdSC$ defined below in Section \ref{general} admit nontrivial coverings.

It is possible to single out in $\AdSC$
a Euclidean manifold $EAdS_d$; we choose the connected real submanifold 
\begin{equation}
EAdS_d=   \{z^{\scriptscriptstyle{E}}= (ix^0, \vec x, x^d)\in \AdSC, \ \    x^d \geq 1 \}  \label{eads}
\end{equation} 
equipped with the Riemannian metric induced by the
scalar product  (\ref{2.1}). 
It is the upper sheet of the two-sheeted hyperboloid
with equation
\begin{equation}
    {x^d}^2 -{x^0}^2- {|\vec{x}|}^2  = 1
\end{equation}
i.e. it is a copy of the Lobachevsky space $H_d$.

Global coordinates may be introduced on $\AdS$ by the map
\begin{equation}
(t,\, \vec{x}) \rightarrow
\left(\sqrt{1+\vec{x}^2}\,\sin t,\ \vec{x},\ \sqrt{1+\vec{x}^2}\,\cos t\right)
\label{2.11}\end{equation}
where $t\in  S_1=\bR/2\pi \bZ$ and $\vec x \in {\bf R}^{d-1}$. Coordinates on $EAdS_d$ are obtained  extending the above map to
$S_1^{(c)}\times \bR^{d-1}$ and taking $t=  is$, namely 
\begin{equation}
(s,\, \vec{x}) \rightarrow
(i\sqrt{1+\vec{x}^2}\,{\rm sh}\  s,\ \vec{x},\ \sqrt{1+\vec{x}^2}\,{\rm
ch}\ s).
\label{parameucl}\end{equation}
Lifting the coordinates 
on the covering $\bR^{d}$ of
$S^1\times \bR^{d-1}$
provides a global coordinate system on $\wt \AdS$.
By further extending the coordinates to 
${\bf C}\times \bR^{d-1}$ provides 
a partial complexification of the covering manifold
$\wt \AdS$;  this complexified manifold contains 
{\em the same} Euclidean space $EAdS_d$ as $\AdSC$.

\section{The null cone and geodesical motion}\label{conepw}

Here we introduce the real null cone of the ambient space (minus its vertex at the origin)
\begin{equation}
\CC_d=\{\xi \in \amb,\ \xi^2={(\xi^0)^2}-{\vec \xi^2}+{(\xi^d)^2}=0,\  \xi\not = 0\}
\end{equation}
and its complexification 
\begin{equation}
\CC_d^{(c)}=\{\zeta= \xi+i\chi \in \ambc,\ (\xi,\xi)-(\chi,\chi)=0, \ (\xi\cdot \chi) =0,\  \zeta\not = 0\}. 
\end{equation}
We single out the following basis manifold  for the  real cone  $\CC_d$:
 \begin{equation}
\gamma_S = \left \{(\xi^{0})^2+({\xi^{d}})^2   = 1 = (\xi^{1})^2+\ldots +({\xi^{d-1}})^2 \right\}.
\end{equation}
The manifold $\gamma_S$ is homeomorphic to $S_1\times S_{d-1}$; it may be parametrized by an angle $\alpha$ and a unit vector on a $(d-1)$-sphere as follows:
\begin{equation}
 \xi(\phi,\vec n)=\left\{\begin{array}{lcl}
\xi^{0} &=&  \sin \phi  \\
\xi^{i} &=&  n^i \ \ \ \ \ , \ \ \ \vec n^2=1   \\
\xi^{d} &=& \cos \phi
\end{array} \right. ;
\label{conecoor}
\end{equation}
it should be also noted that both $\xi$ and $-\xi$ belong to $\gamma_S$ and \begin{equation}\xi(\phi+\pi,\vec n)=-\xi(\phi,-\vec n).\end{equation}
In the special two-dimensional   $\gamma_S$ is the disjoint union of two disconnected components 
\begin{equation}
\gamma_{S(2)}^{\pm} = \left \{(\xi^{0})^2+({\xi^{d}})^2   = 1, \ \ \  \xi^{1}= \pm 1 \right\}.\label{basis2}
\end{equation}
Similarly $\CC_d^{(c)}$ admits the basis manifold
 \begin{equation}
\gamma_S^{(c)} = \left \{(\zeta^{0})^2+({\zeta^{d}})^2   = 1 = (\zeta^{1})^2+\ldots +({\zeta^{d-1}})^2 \right\}
\end{equation}
which may be coordinatized by complexifying the parameters in Eq. (\ref{conecoor}). 
\subsection{Geodesics}

\begin{figure}[ht]
  \centering
  \includegraphics[width=0.7\textwidth]{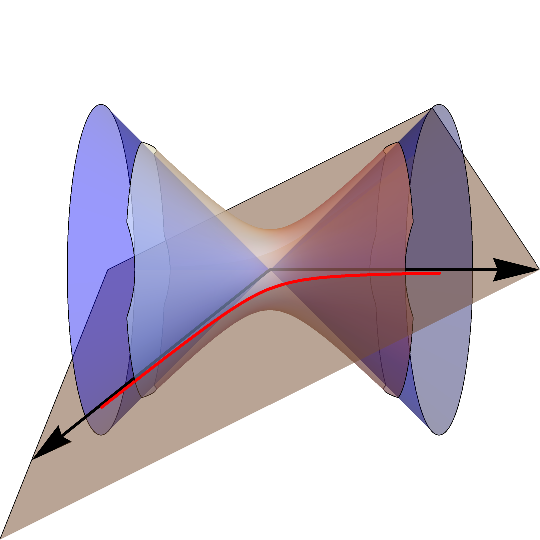}
  \caption{Spacelike geodesics are parametrized by two real null vectors in the ambient space; they identify the asymptotic directions of the geodesics at spacelike infinity (the boundary).}
  \label{fig:geo}
\end{figure}

The study of geodesics gives a clue for the construction and the interpretation of AdS  plane waves suitable for  quantum field theory. 
Since the AdS manifold is a real section of a complex sphere,   geodesics of the AdS spacetime  are conic sections of $AdS_d$ by 2-planes containing the origin of
the ambient space $\amb$ (see Figure \ref{fig:geo}; see also \cite{cacc}). 

For spacelike geodesics the 2-planes intersect the real null cone $\CC_d$
along two generatrices.
Let  $\xi$ and $\eta$  two null vectors identifying the generatrices, chosen in such a way that $(\xi, \eta) >0$ (changing $\eta$ with $-\eta$ if necessary);   a parametrization of the geodesic in terms of the proper length  $s$ is then constructed as follows:
\begin{equation}
x^\mu(s) = \frac{\xi^\mu e^{\alpha s} + \eta^\mu e^{-\alpha s}}{\sqrt{2\xi \cdot  \eta}}
\end{equation}
where the constant $\alpha^2 = -\dot x^2>0$.
Real null vectors of the ambient space $\amb$ may thus be interpreted as indicating asymptotic spacelike momentum directions on the AdS manifold.

On the other hand, planes containing timelike geodesics do not intersect the real cone  $\CC_d$ but they do intersect its complexification 
\begin{equation}
\CC_d^{(c)}=\{\zeta= \xi+i\chi \in \ambc,\ (\xi,\xi)-(\chi,\chi)=0, \ (\xi\cdot \chi) =0,\  \zeta\not = 0\} 
\end{equation}
along two complex conjugated directions. 
Timelike geodesics on the real manifold $\AdS$ can be thus parameterized by one complex null vector as follows:
\begin{eqnarray}
x^\mu(\tau) &=& \frac{\zeta^\mu e^{i \beta \tau} + {\zeta^\mu}^* e^{-i \beta  \tau}}{\sqrt{2\zeta \cdot \zeta^*}}, \ \ \ (\zeta\cdot \zeta)=0 \label{geoc}\\ 
v^\mu(\tau)& = & i \beta \frac{\zeta^\mu e^{i \beta \tau} - {\zeta^\mu}^* e^{-i \beta  \tau}}{\sqrt{2\zeta \cdot \zeta^*}}, \ \ \ (v\cdot v)=\beta^2.
\end{eqnarray}
In terms of $\zeta$ we may construct a complete set of  conserved quantities that characterise the geodesic motion:
\begin{equation}
K^{\mu\nu} = \frac{m (x^\mu(\tau)v^\nu (\tau) -x^\nu(\tau) v^\mu (\tau))}{\sqrt{ v\cdot v}} =\frac{  i \, m (\zeta^\mu  {\zeta^\nu}^* -\zeta^\nu  {\zeta^\mu}^*) }{ \zeta \cdot  \zeta^*} =  \frac{i\, m( \zeta\wedge   {\zeta}^* )^{\mu\nu}}{ \zeta \cdot \zeta^*} 
\end{equation}
where  $m$ is a mass parameter of the point particle in geodesical motion.

The parametrization (\ref{geoc}) demonstrates that every temporal geodesic passing through a given event 
$x$
 also passes through its antipodal event 
$-x$. The proper time interval between 
$x$
 and 
$-x$ along each of these geodesics is 
$\pi R$ (where we have set 
$R
=
1$ in the formulae), which corresponds to half of a "Great Era,"  \cite{gibbons}.

On the other hand, there are events that are not connected by an arc of geodesics and therefore $\AdS$ is not geodesically convex.

These characteristics also manifest in the universal covering space  $\widetilde\AdS$; In this covering manifold, geodesics continue to refocus at intervals of  $\pi R $ i.e. half of the period. Thus, the intrinsic time periodicity; the time periodicity of the AdS spacetime does is not eliminated by passing to the cover, a point rightly emphasized by G. Gibbons \cite{gibbons}.

\section{The chiral tuboids $\ZZ_{+}$ and $\ZZ_{-}$} \label{general}

Let $\GG$ be the Lie algebra of $G_0$ 
and let $M_{0d}$ the generator of rotations in the $(0,d)$-plane so that
\begin{equation}
e^{tM_{0d}} = \left (
\begin{array}{ccc}
\cos t & \ldots & \sin t\\
\vdots & 1 & \vdots \\
-\sin t & \ldots & \cos t
\end{array}
\right ).\ 
\label{2.4}\end{equation}
The following  subsets of $\GG$ plays a central role in our construction:
\begin{eqnarray}
&& C_0=\{M\in \GG: \ M= g\, M_{0d}\,g^{-1}\ :\ g \in G_0\},  \cr 
&& C_+ =\bigcup_{\rho>0}\rho C_0. \label{c1}
\label{2.8}
\end{eqnarray} 
 Let us now recall the definition  \cite{bem} of the forward and backward {\em chiral} tuboids by the following definitions:
\begin{equation}
\ZZ_{\pm} = \{\exp(\tau M)\,x\ :\ M \in C_0,\ \ x \in X_d,\ \ 
\tau \in \bC, \ \Im \tau \gtrless  0\ \}.  
\label{3.1.1}\end{equation}
Their universal coverings  $\widetilde \ZZ_{\pm}$  are given by the same formulae  
where $\exp(\tau M)$ is 
understood in the sense of the covering as an element of $\wt G_0^{(c)}$
and $x$ as an arbitrary element of $\wt \AdS$.
The chiral tuboids are manifestly invariant under the action of $G_0$. 
They can be equivalently described as follows  \cite{bem}:
\begin{eqnarray}
 \ZZ_{\pm} = \{z=x+iy \in \ambc\ :  \ \scalar{y}{y}  >0,\  \epsilon(z) \gtrless 0  \},
\label{3.2-}
\end{eqnarray} 
where  we denoted
\begin{equation}
 \epsilon(z) = y^0 x^d - x^0 y^d.
\label{3.1.0}\end{equation}
The identity  
\begin{eqnarray}
&& \epsilon(z)^2= (\scalar{y}{y}) (1+|\vec{x}|^2+| \vec{y}|^2) +(\scalar{y}{y}) ^2+ |\vec{y}|^2  + |\vec{x}|^2 |\vec{y}|^2 - (\vec{x}\cdot \vec{y})^2 \label{2.5.gg}
\end{eqnarray}
that is valid for any  complex event  $z\in \AdSC$, implies that  $\epsilon(z)^2$  cannot vanish when $\scalar{y}{y} >0$; Eq. (\ref{3.2-}) then  shows that 
$\ZZ_{+}$ 
and  
$\ZZ_{-}$ are  disjoint open  
subsets of $\AdSC$ while Eq.  (\ref{3.1.1}) shows that they are 
connected.
$\ZZ_{+}$  and  $\ZZ_{-}$ may be generated simply by acting on the base point $e_d$ by a complex rotation followed by a Lorentz transformation:
 \begin{eqnarray}
\ZZ_{\pm} = \{g\, \exp(it
M_{0d})\,e_d\ :\
g \in G_0,\ \ \ t \gtrless 0\}.
\label{3.2.1zero} 
 \end{eqnarray}
 If  $z$  belongs to  either of the closed tuboids $\ovl{\ZZ}_{\pm }$     then  for any  $M\in C_0$  and $ \Im \tau \gtrless 0 $ 
the point $ \exp(\tau M)z$ belongs to either of their interiors $\ZZ_{\pm} $ . 

\vskip 10 pt

\noindent The following geometrical properties are of central importance for AdS quantum field theory \cite{bem}:
 \begin{description}
 \item{\it (i) } The Euclidean AdS manifold $EAdS_d$ is contained in $\ZZ_+\cup\ZZ_-\cup \AdS$.
\item {\it (ii)} The image of either of the domains $\ZZ_{+}$  or $\ZZ_{-}$
under the coordinate map $z \rightarrow z^d = \scalar{e_d}{z}$ is the cut-plane
$\Delta = {\bf C} \setminus[-1,1]$.
\item {\it (iii)} 
The image of the domain $\ZZ_{-} \times \ZZ_{+}$ (or $\ZZ_{+} \times
\ZZ_{-}$) by the scalar product
mapping $z_1,z_2 \rightarrow \scalar{z_1}{z_2}$ is the cut-plane
$\Delta$.
\item  {\it (iv)}
The map $z_1, z_2 \rightarrow \scalar{z_1}{z_2}$ of $\ZZ_{-} \times \ZZ_{+}$
onto $\Delta$ can be lifted to a map of
$\widetilde\ZZ_{-} \times \widetilde\ZZ_{+}$ onto
the covering $\widetilde\Delta$ of the cut-plane
$\Delta$.
\end{description}

\vskip5pt
A more explicit characterization of the tuboids may be obtained by using the global complex  coordinates of the complex AdS manifold that we write here as follows: 
\begin{equation}\label{AdSCParam} z(t+is,\psi+i \phi, \vec a+i \vec b)=  \left\{  \begin{array}{l}\cosh(\psi+i \phi)\sin (t+i s), \cr \sinh(\psi+i \phi)\ (\vec  a+i \vec b), \cr \cosh(\psi+i \phi)\cos (t+i s), \end{array} \right. \ \ \ 
|\vec a|^2-|\vec b|^2=1, \ \ \vec a \cdot \vec b = 0.
\end{equation}
The real covering manifold $\wt \AdS$ is coordinatized by taking $\phi=0$, $s=0$, $\vec b=0$ and by unfolding the $t$ coordinate.

The chiral tuboids $\ZZ_{-}$ 
and  
$\ZZ_{+}$ are semi-tubes:  they are domains  in ${\bf C}^d$ 
invariant in one real direction:
\begin{eqnarray}
&& {\ZZ_+}\ : \ \ \ \ \sinh s >  \sqrt{ \frac{|\vec b|^2 \, \sinh ^2(\psi )+\left(|\vec b|^2+1\right) \sin ^2(\phi )}{\sinh ^2(\psi )+\cos
   ^2(\phi )}}; \\
  &&  {\ZZ_-}\ : \ \ \ \ \sinh s <-  \sqrt{ \frac{|\vec b|^2 \, \sinh ^2(\psi )+\left(|\vec b|^2+1\right) \sin ^2(\phi )}{\sinh ^2(\psi )+\cos
   ^2(\phi )}}.
\end{eqnarray}
bordered by the surfaces $s = s_\pm(\psi, \phi,|\vec b^2|)$.  $\widetilde \ZZ_{\pm}$  are given by the same formulae  
where the real variable $t$ has been unfolded.

\section{The analytic structure of two-point functions}

\label{axiom}

The geometrical properties discussed in the previous section allow to fully characterize  the two-point functions of scalar fields on either $\wt \AdS$ or on the uncovered manifold $AdS_d$ for integer values of the mass parameters. We assume the following general properties \cite{bem}: 
\begin{description}
\item{1.} {\bf Covariance:} ${\cal W}$ is invariant under
the action of the group $\wt G_0$ (resp. $ G_0$), i.e. 
\begin{equation}
 {\cal W}(g x_1, g x_2)  =
 {\cal W}( x_1,  x_2 )
\label{4.3}
\end{equation}
for all $g\in \wt G_0$ (resp. $ G_0$).

\item{2.} {\bf Locality:} if $x_{1}$ and $x_{2}$ are space-like separated then
the commutator  \begin{equation}
{\cal C}({x_{1}},x_{2})= {\cal W}({x_{1}},x_{2})
-{\cal W}({x_{2}},x_{1})
\label{4.4}\end{equation}vanishes.

\item{3.} {\bf Positive Definiteness:} For any smooth test function $f$ in a suitable test-function space $\TT(\wt\AdS)$  the following positivity property holds:
\begin{equation}
\WW(\overline f f) = \int  {\cal W}( x_1, x_2) \overline {f(x)} f(y)  d\sigma(x)d \sigma(y) \geq 0.
 \label{4.5}\end{equation}
\end{description}
A reconstruction procedure essentially identical to that 
of relativistic quantum field theory  \cite{pct} then provides a Fock-Hilbert space $\HH$, a continuous unitary
representation $g \rightarrow U(g)$ of $\wt G_0$ (resp. $
G_0$); a unit vector $\Omega \in \HH$, invariant
under $U$ and a representation of the field as an operator valued distribution $f \rightarrow \varphi({f})$  such that 
\begin{eqnarray}\WW(fg) = (\Omega,\ \varphi(f)\varphi(g)\Omega)  \  \ for \  all\ \ 
f,g \in \TT 
\end{eqnarray}
and that
\begin{equation}
U(g)\,\varphi(f)\,U(g)^{-1} = \varphi(f_{\{g\}}),\ \ \ \ f_{\{g\}}(x)= f(g^{-1}x), \ \ \ \ 
U(g)\,\Omega = \Omega\ 
\label{4.7}\end{equation}
for all $g \in \wt G_0$ (resp. $G_0$).

\vskip 10 pt

Now we need a physical criterion to select among all possible quantizations realizing the above conditions. Operators representing elements of $C_+$ (see Eq. \ref{c1}) may be interpreted as Hamiltonian operators; a spectral condition selecting a class of relevant quantization amounts to assuming positivity of the spectra of such operators. More precisely,  to every element $M\in C_+$ 
we can associate the one-parameter subgroup $t \rightarrow \exp tM$ of
$\wt G_0$ (resp. $ G_0$) and 
a self-adjoint operator $\wh M$ acting in
$\HH$ such that $\exp it\wh M = U(\exp tM)$ for all $t \in \bR$.
With these notations, we assume the following

\begin{description}
\item{4.} {\bf (Strong) Spectral Condition:}
For every $M \in C_+$, every $\Psi \in \HH$, and every $\CC^\infty$
function $\wt f$ with compact support contained in
$(-\infty,\ 0)$,
\begin{equation}
\int_{\bR} \left ( \int \wt f(p)\,e^{-itp}\,dp \right )\,
U(\exp tM)\,\Psi\,dt = 0\ .
\label{4.8}\end{equation}
Equivalently $\wh M$ has its spectrum contained in $\Rp$.
\end{description}

The strong spectral condition  actually broadens to cover general (interacting) QFT's whose   truncated $n$-point functions do not vanish  \cite{bem}. When restricted to the case of two-point functions 
 the  condition tells us that the 
two-point function is not just any distribution on $\wt \AdS\times \wt \AdS$  but is realized as a boundary value of a certain analytic function: \cite{bem}:
\begin{description}
\item{$4^\prime$.} { \bf Normal analyticity  for two-point functions:}
{\em $\mathcal W\left( x_1,x_2\right)$
is the boundary value of a  function $ W\left(
z_1, z_2\right) $ holomorphic in the  domain
$\wt{\cal Z}_{-}\times \wt{\cal Z}_{+}$} 
\begin{align}
&\WW(x_1, x_2) = (\Omega,\ \varphi(x_1)\varphi(x_2)\,\Omega) =
\lim_{\stackrel{
 \wt{\cal Z}_{-}\ni \  z_1 \to x_1}{ \scriptscriptstyle{\wt{\cal Z}_{+} \ni \ z_2 \rightarrow x_2}}}
W(z_1,z_2).  \label{spp}
\end{align}
\end{description}
AdS invariance then implies that to 
$W(z_1,z_2)$ there corresponds a function 
of a single complex variable  that can be identified with the scalar product 
$z_1\cdot z_2$ when  $z_1\in \ZZ_-$ and $z_2\in \ZZ_+$;  the function $  W(z_1\cdot z_2) $ is called {\em the reduced two-point function}.
Complex AdS invariance and normal analyticity   then imply the following 

\begin{description}
\item {$4^{\prime\prime}$.} {\bf  Maximal analyticity property}:
the reduced two-point function {\em $W(z_1\cdot z_2)$ extends to a function
analytic in the  covering $\wt\Delta_1$ of the
cut-plane 
$\Delta_1 = \{{\bf C} \setminus[-1,1]\}. 
$ 
For theories single-valued  on $AdS_d$, $W(z_1\cdot z_2)$ is  analytic in $\Delta_1$.} 
\end{description}

Note that the two-point function of any field
satisfying locality AdS invariance and the spectral condition
enjoys  {\em maximal analyticity}, 
as it happens in the Minkowski \cite{pct}
and de Sitter  cases \cite{bgm,bm}.   Importantly, this property holds even in the absence of the positive definiteness condition. However, if positive definiteness is satisfied, then the theory admits a quantum mechanical Hilbert space interpretation, allowing for a consistent quantum probabilistic interpretation.

\section{Klein-Gordon  equation, homogeneous functions  and plane waves}
\label{dal}

Let us begin by recalling the relationship between the Laplace-Beltrami operator $\Box_{g} = g_{\mu\nu} \partial^\mu \partial^\nu$, defined with respect to the metric $g_{\mu\nu}$ of the ambient space $\amb$, and the anti-de Sitter d'Alembert operator, which we will denote simply as $\Box$.

To this end let $\dil$ be the infinitesimal generator of dilations so that  
$
(\dil f)(x) = x^\mu \partial_\mu f(x)
$ and let $ \MM_{\mu\nu} =    ( x_{\mu}\partial _{\nu} -x_{\nu}\partial _{\mu})$. The following relation holds on $\amb$:
\begin{equation}
\MM^2 = \MM^{\mu\nu}  \MM_{\mu\nu}= 2x^2\Box_{g} -2(d-1)\dil -2 \dil^2.
\label{f.8}\end{equation}
If  $f$ and $h$ are smooth functions on $\amb$
and $h$ is $G_0$-invariant, then
\begin{equation}
\MM_{\mu\nu} (h(x)f(x)) = h(x) \MM_{\mu\nu} f(x)\ .
\label{f.9}\end{equation}
This remains true if $h$ is a $G_0$-invariant distribution
such as $\delta((x\cdot x) -1)$.
In particular the operators  $\MM_{\mu\nu}$ and $\MM^2$ commute with
the restriction to invariant submanifolds of $\amb$ and  also
with dilations.

If $f$ is a smooth function or a distribution
homogeneous of degree zero
on the open set
$
\Omega=\{x \in \amb\ :\ (x,x) > 0\}
$
then
\begin{equation}
\Box (f|_{\AdS}) = (\Box_{g} f)|_{\AdS} = \frac 12 \MM^2 f |_{\AdS}
\end{equation}
because a homogeneous function of  degree zero 
 is dilation invariant. 
Since any smooth function or any distribution defined on $AdS_d$ has a unique  homogeneous extension 
of degree zero in the open set $\Omega$, it follows that \begin{equation}
\Box \phi = \frac 12  \MM^2\phi
\label{f.10}\end{equation}
for every smooth function or distribution $\phi$  defined on $AdS_d$. 

Suppose now that $f$, defined in the ambient space $\amb$
and {\em homogeneous of degree $\lambda$},  solves the massless equation
$\Box_{g}f = 0$. From Eq. (\ref{f.8})  we get that
\begin{equation}
\MM^2 f(x) = -2\lambda(\lambda+d-1) f(x).
\label{f.11}\end{equation}
 Eqs. (\ref{f.10}) and (\ref{f.11}) then  imply that  the restriction of $f$ to the anti-de Sitter manifold $\AdS$ satisfies  the Klein-Gordon equation
\begin{equation}
(\Box + m^2_\lambda) (f(x)|_{\AdS}) = 0
\label{cdskg}\end{equation}  
with {\em complex} squared mass
\begin{equation}
m^2_\lambda =  \lambda(\lambda+d-1). \label{mass0}
\end{equation}
Plane waves provide the most important example of homogeneous solutions of the massless equation in the ambient space:
\begin{equation}
x,\zeta \rightarrow (\scalar{x}{\zeta})^\lambda ,
\label{pwholo}
\end{equation}
where $\lambda$ is a complex parameter and $\zeta$ is a complex null vector in $\CC_d^{(c)}$. The restriction $\phi_\lambda(x,\zeta)$ of (\ref{pwholo}) to the anti–de Sitter manifold satisfies the massive Klein–Gordon equation (\ref{cdskg}):
\begin{equation}
\phi_\lambda(x,\zeta) = (\scalar{x}{\zeta})^\lambda, \qquad x\in \AdS .
\label{ww}
\end{equation}
The invariance of $m_\lambda^2$ under the involution
\begin{equation}
\lambda \to \bar{\lambda}=  1-d-\lambda
\label{involution}
\end{equation}
implies that, for any given complex null vector $\zeta$, there exists a second solution of Eq. (\ref{cdskg}), obtained by applying this involution to the exponent of the wave:
\begin{equation}
\left(\Box + m_\lambda^2\right)\phi_{1-d-\lambda}(x,\zeta)=0 .
\label{p.2abis}
\end{equation}
To establish a connection with notations commonly used in the AdS--CFT literature, we introduce three additional complex parameters: $\nu$, $\Delta_-$, and $\Delta_+$. These parameters are related to the homogeneity degree $\lambda$ through
\begin{equation}
\lambda = -\frac{d-1}{2} + \nu = -\Delta_- .
\label{nnu}
\end{equation}
Under the involution (\ref{involution}) 
\begin{equation}
\bar{\lambda} = 1-d-\lambda = -\frac{d-1}{2} - \nu = -\Delta_+  \label{num}
\end{equation}
the complex parameter $\nu$ simply changes sign,
\begin{equation}
\nu \rightarrow  -\nu ;
\label{invonu}
\end{equation}
the involution does not coincide with complex conjugation unless $\nu$ is purely imaginary, which corresponds to a tachyonic wave devoid of physical interpretation in AdS.

We may further clarify the relation between the homogeneity degree $\lambda$ and the scaling dimnsion  $\Delta_-$  by using  the Poincar\'e coordinates (\ref{coordinates}) and (\ref{conecoor0}) to parametrize the plane wave:
\begin{equation}
(x(t,u) \cdot \xi(\eta))^\lambda=(x(t,u) \cdot \xi(\eta))^{-\frac{d-1}2+\nu}=
 \left(\frac{u}{2}-\frac{\left({\x}-{\eta}\right)^2}{2u}\right)^{-\Delta_-} \simeq \frac{ u^{\Delta_-}}{(\left({\x}-{\eta}\right)^2)^{\Delta_-}}
 \end{equation}
 \begin{equation}
(x(t,u) \cdot \xi(\eta))^{\bar\lambda}=(x(t,u) \cdot \xi(\eta))^{-\frac{d-1}2-\nu}=
 \left(\frac{u}{2}-\frac{\left({\x}-{\eta}\right)^2}{2u}\right)^{-\Delta_+} \simeq \frac{ u^{\Delta_+}}{(\left({\x}-{\eta}\right)^2)^{\Delta_+}}
 \end{equation}
(unless $\x$ and $\eta$ are colinear). 
The squared mass can then be written as
\begin{equation}
m_\lambda^2 = \nu^2 - \frac{(d-1)^2}{4} = -\Delta_+ \Delta_- .
\label{massnu}
\end{equation}
Let us assume that the spacetime dimension parameter $d$ is real. For real $\lambda$ (equivalently, for real $\nu$), the squared mass (\ref{massnu}) reaches its minimum negative value at $\nu = 0$:
\begin{equation}
m^2_{\min} = -\frac{(d-1)^2}{4}.
\label{masmin0}
\end{equation}
This minimum corresponds to the homogeneity degree
\begin{equation}
\lambda_d = -\frac{d-1}{2}.
\label{masmin1}
\end{equation}
In the following, we will use $\lambda$ and $\nu$ interchangeably.
\vskip 5 pt

   The main challenge now emerges: although the complex parameter $\lambda$ in the above definition can take any value, the plane waves themselves are only locally well-defined on AdS$_d$. An exception occurs when $\lambda$ is an integer; in this case, the plane waves are globally well-defined. For non-integer $\lambda$, however, constructing single-valued plane waves requires additional effort on the covering space $\widetilde{\text{AdS}}_d$. This issue will be discussed in detail in the following section.

\vskip 10 pt
\section{Chiral cones} \label{cone}
The heuristic definition (\ref{pwholo}) is formally identical to the corresponding one employed in the de Sitter case \cite{bgm,bm}; however, the analytic structure in the present context is significantly more intricate and requires a more elaborate analysis to be applicable for real non-integer and, more generally, for complex $\lambda$. With this in mind, we now introduce the \textit{chiral cones}, defined as the following two $G_0$-invariant subsets of $\CC_d^{(c)}$:  
\begin{equation}
\CC_{\pm} = \{\exp(\tau M)\, \xi \ :\ M \in C_0,\ \ \xi\in \CC_d\ \ \
\tau \in \bC, \  \ \Im \tau \gtrless0\}. 
\label{3.1.00}\end{equation}
They  may also  be  identified with the following subsets of $\CC_d^{(c)}$ 
\begin{eqnarray} \label{chiralcones+2}
 \CC_{\pm} = \{\zeta=\xi+i\chi \in \ambc:\
\scalar{\xi}{\xi}  = \scalar{\chi}{\chi}  >0, \ \scalar{\xi}{\chi}  = 0,  \  \epsilon(\zeta)\gtrless 0\}
\end{eqnarray}
and  may be generated simply by acting on the  point $e_{d-1}+e_d$ as follows (see Appendix \ref{Appa}):
 \begin{eqnarray}
\CC_{\pm} = \{g\, \exp(it
M_{0d})(e_{d-1}+e_d)\ :\
g \in G_0,\ \ \ t \gtrless 0\}.
\label{3.2.1ab} 
 \end{eqnarray}
Here we summarize the properties that make  use of plane waves on the AdS manifold and  its covering possible:  
\begin{enumerate}
 
\item
Let \begin{eqnarray} \label{chiralcones+3}
\ovl\CC_{+} = \{\zeta\in \CC_d^{(c)}:\,
\scalar{\chi}{\chi}  \geq 0,  \,  \epsilon(\zeta) \geq 0,\, \zeta\not=0\},\\
\ovl\CC_{-} = \{\zeta\in \CC_d^{(c)}:\,
\scalar{\chi}{\chi}  \geq 0,  \,  \epsilon(\zeta) \leq 0,\, \zeta\not=0\}.
\end{eqnarray}
be the punctured closures of the tubes (including their real boundary points, minus the origin). It is obvious that  
\begin{equation}
\exp(\tau M)( \ovl{\CC}_{\pm} )\subset  \CC_{\pm} ,\ \ \  \makebox{for}\ \ M\in C_0, \ \ \Im \tau \gtrless 0. \label{420}
\end{equation}

\item

The image of either $\CC_{+}$  or $\CC_{-}$
under the coordinate map $\zeta \rightarrow \zeta^d = \scalar{e_d}{\zeta}$ is the punctured complex-plane
${\bf C} ^* = {\bf C} \setminus\{0\}$. 
This fact is intuitively clear  as the chiral cones are  the limit of the chiral tuboids when the radius $R$ tends to zero. 

An explicit check is also easy: using Eq. (\ref{3.2.1}) for  $ \zeta \in \CC_{+}$
\begin{equation}
\scalar{e_d}{\zeta} = x\cdot \exp(it M_{0d})(e_{d-1}+e_d), \label{423} 
\end{equation}
where  $x =g^{-1} e_d= \exp(s M_{0d})\,(0,\ \vec{x},\ \sqrt{1+\vec{x}^2})$, \  $s= \arctan (x^0/x^d)$ and $t > 0$  ; therefore 
\begin{equation}
\scalar{e_d}{\zeta} = - x^{d-1} +\sqrt{1+{\vv x}^2}\,\cos(it -s) \in {\bf C}^*. \label{09o}
\end{equation}
Conversely, any $w \in {\bf C}^*$ can be written as
$w = \lambda \cos(u+iv)$, with $v>0$ and $\lambda>0$, i.e. $w = \scalar{e_d}{(\lambda \exp((u+iv)M_{0d})(e_{d-1}+e_d))}$
is in the image of $\CC_{+}$.

\item

Let us consider  now  $z \in \ZZ_{-}$, $\zeta \in\ovl{\CC}_{+}$;
there are   $g \in G_0$,  $M \in C_0$ and a complex $\tau$   with  $\Im \tau >0$ such that 
\begin{equation}
\scalar{z}{\zeta} = \scalar{e_d}{\exp(\tau M) g ^{-1} \zeta }
\end{equation}
Eqs. (\ref{420}) and (\ref{09o}) then imply that 
the image  of $\ZZ_{-} \times \ovl{\CC}_{+}$   by the scalar product
mapping $z,\zeta \rightarrow \scalar{z}{\zeta}$ is also the punctured complex-plane  ${\bf C} ^*$ and the same is obviously true for $\ZZ_{+} \times
\ovl{\CC}_{-}$.
In words: the scalar product $\scalar{z}{\zeta}$ 
cannot vanish when either $z\in \ZZ_{-} $ and $ \zeta \in \ovl{\CC}_{+}$ or when $z\in \ZZ_{+} $ and $ \zeta \in \ovl{\CC}_{-}$.

\item

The map $z,\zeta  \rightarrow \scalar{z}{\zeta}$ of $\ZZ_{\pm} \times \ovl{\CC}_{\mp}$
onto ${\bf C} ^*$ can be lifted to a map of
$\widetilde\ZZ_{\pm} \times \ovl{\CC}_{\mp}$ onto
the covering $\widetilde{\bf C}^*$ of the punctured plane ${\bf C} ^*$. by abuse of language we denote the lifted scalar product by the same symbol.

\item

Finally, let us consider  $\chi \in \CC_{-},\ \zeta \in\ovl{\CC}_{+}$, ;
there are   $g \in G_0$,  $M \in C_0$ and a complex $\tau$   with  $\Im \tau >0$ such that 
\begin{equation}
\scalar{\chi}{\zeta} = \scalar{(e_{d-1}+e_d)}{ \zeta' }, \ \ \ \zeta' = \exp(\tau M) g^{-1} \zeta \in \CC_+
\end{equation}
We are thus led to examine the scalar product 
\begin{equation}
\scalar{\xi}{\exp(it M_{0d}) (e_{d-1}+e_d)}= i\  \sinh(t)\,\xi^0 -\xi^{d-1}+ \cosh(t)\xi^d \ \  \xi \in \CC_d, \ \  t>0
\end{equation}
and this can never be zero.

In conclusion the scalar product $\scalar{\chi}{\zeta}$ 
cannot vanish when either $\chi\in \CC_{-} $ and $ \zeta \in \ovl{\CC}_{+}$ or when  $ \chi \in \ovl{\CC}_{-}$ and $\zeta\in \CC_{+} $ and.
 \end{enumerate}

\subsection{Plane waves again}
We are now in position to give a global definition of the anti-de Sitter plane waves. The same problem has been solved long ago in the de Sitter case \cite{bgm,bm} (see Appendix \ref{aapb}). Given  any complex number $\lambda$ the natural domains of definition for plane waves in the AdS universe are    $\ZZ_+ \times \ovl \CC_-$ and $\ZZ_- \times \ovl \CC_+$: 
\begin{equation}
\wt \ZZ_\pm \times \ovl \CC_\mp \ni \zeta,z \to \phi^{\pm}_\lambda(z,\zeta) = 
\left({ z\cdot \zeta}\right)^{\lambda}  =  e^{\lambda \log \left({ z\cdot \zeta}\right) }.   \label{waves}
\end{equation}
A previous attempt \cite{pene1} to introduce global waves in AdS was based on a local $i \epsilon$-prescription. Here, in their global domains of definition, the waves are univalued, holomorphic  solutions of  the  Klein-Gordon equation  with complex  squared mass
$
    m^2_\lambda = \lambda(\lambda+d-1).$ 
    
   In the following we will also need to consider plane waves in subsets of $\ZZ_+ \times \ovl \CC_+$ and $\ZZ_- \times \ovl \CC_-$ bordered; in that case it will be necessary not to trespass the  hyperplanes where the argument of the waves vanishes.  Hence, the admissible domains in these bordered subsets are restricted by these hyperplanes, which act as natural boundaries ensuring the holomorphicity and single-valuedness of the plane wave solutions. This condition secures the integrity of the analytic continuation of the waves and preserves the properties necessary for their role  in the anti-de Sitter spacetime framework.

\section{Klein-Gordon fields}
\label{kgf}
Klein--Gordon fields provide the simplest example of the analytic structure described in section \ref{axiom}. 
Their two-point functions are distributions on $\widetilde{\AdS}\times\widetilde{\AdS}$ satisfying the AdS Klein--Gordon equation with respect to both variables:
\begin{equation}
\left(\square_{x_1}+m_\lambda^2\right)\mathcal{W}(x_1,x_2)
=
\left(\square_{x_2}+m_\lambda^2\right)\mathcal{W}(x_1,x_2)=0,
\label{kg}
\end{equation}
where the (complex) squared mass is parametrized by the complex number $\lambda$ as in eq.~(\ref{mass0}).

For any such $\lambda$ there exist two invariant maximally analytic solutions satisfying all the properties listed in section \ref{axiom}, except possibly positive definiteness (unitarity). They are function of the complex variable $z=z_1 \cdot z_2$ and  are mapped into each other by the involution (\ref{involution}): 
\begin{align}
W^{(d)}_\lambda(z_1,z_2)
&= W^{(d)}_\lambda(z)  
= \frac{e^{-i\pi \frac{d-2}2}}{(2\pi)^{\frac d 2}}
\big(z^2-1\big)^{-\frac{d-2}4}
Q^{\,\frac{d-2}{2}}_{\frac{d-2}{2}+\lambda}(z)
\label{kgtp}
\\
&= \frac{z^{1-d-\lambda}
\Gamma(d-1+\lambda)\,
{}_2F_1\!\left(\frac{d-1+\lambda}{2},\frac{d+\lambda}{2};
\frac{d+1+2\lambda}{2};z^{-2}\right)}
{2^{d+\lambda}\pi^{(d-1)/2}
\Gamma\!\left(\frac{d+1}{2}+\lambda\right)}
\label{kgtp1}
\\
W^{(d)}_{1-d-\lambda}(z_1,z_2)
&=
\frac{e^{-i\pi \frac{d-2}2}}{(2\pi)^{\frac d 2}}
\big(z^2-1\big)^{-\frac{d-2}4}
Q^{\,\frac{d-2}{2}}_{-\frac{d}{2}-\lambda}(z).
\label{kgtp2}
\end{align}
By using the parameter $\nu$ introduced in eq.~(\ref{nnu}), the above solutions can also be written as
\begin{align}
W^{(d)}_{\nu(\lambda)}(z_1,z_2)
&=
\frac{e^{-i\pi \frac{d-2}2}}{(2\pi)^{\frac d 2}}
\big(z^2-1\big)^{-\frac{d-2}4}
Q^{\,\frac{d-2}{2}}_{-\frac12+\nu}(z)
\label{kgtpnu}
\\
&=
\frac{\Gamma\!\left(\frac{d-1}{2}+\nu\right)
(2 z)^{-\frac{d-1}{2}-\nu}
\,{}_2F_1\!\left(\frac{d-1}{4}+\frac{\nu}{2},
\frac{d+1}{4}+\frac{\nu}{2};
\nu+1;
\frac{1}{z^2}\right)}
{2\pi^{\frac{d-1}{2}}\Gamma(\nu+1)} ,
\label{kgtp1nu}
\\
W^{(d)}_{-\nu(\lambda)}(z_1,z_2)
&=
\frac{e^{-i\pi \frac{d-2}2}}{(2\pi)^{\frac d 2}}
\big(z^2-1\big)^{-\frac{d-2}4}
Q^{\,\frac{d-2}{2}}_{-\frac12-\nu}(z).
\label{kgtp2nu}
\end{align}
As expected, the involution acts simply by changing the sign of the parameter $\nu$.

In the above formulae, and in what follows, $P^\alpha_\beta$ and $Q^\alpha_\beta$ denote the \emph{associated Legendre functions}, which are meromorphic in their indices \cite{Bateman}. As functions of the complex variable $z$, they are analytic in the cut plane
\begin{equation}
\Delta_2=\mathbf{C}\setminus[-\infty,1].
\end{equation}
As explained in Section~\ref{general}, the AdS scalar product $z_1\!\cdot\! z_2$ lifts to a map
\[
\widetilde{\mathcal Z}_{-}\times\widetilde{\mathcal Z}_{+}\to \widetilde{\Delta}_1,
\]
(the covering of the cut plane
$
\Delta_1=\mathbb{C}\setminus[-1,1]).
$
Equations (\ref{kgtp})--(\ref{kgtp2nu}) are to be understood in this sense.

The Legendre function of the second kind 
$Q^{\,\frac{d-2}{2}}_{-\frac12+\nu}(z)$ 
(or equivalently $Q^{-\frac{d-2}{2}}_{-\frac12+\nu}(z)$), 
multiplied by the factor $(z-1)^{-\frac{d-2}{4}}(z+1)^{-\frac{d-2}{4}}$, 
are holomorphic in $\wt{\Delta}_1$, a domain naturally adapted to the anti-de Sitter  geometry. 
Points on the cut $[-1,1]$ of the fundamental sheet $\Delta_1$ correspond to pairs of timelike separated events on ${\AdS}$.

In contrast, the following \emph{Legendre functions of the first kind}\footnote{In this case the sign of the upper index is uniquely fixed; these functions are also known as Gegenbauer functions \cite{Bateman}.}, multiplied by the same prefactor 
\begin{equation}
(z-1)^{-\frac{d-2}{4}}(z+1)^{-\frac{d-2}{4}}
P^{-\frac{d-2}{2}}_{-\frac12+i\kk}(z),
\qquad \kk\in\mathbb{C},
\end{equation}
possess analyticity properties that are naturally adapted to the Lorentzian geometry of the de Sitter manifold \cite{bgm,bm} (see Appendix \ref{aapb}) as 
they are analytic in the cut plane
\begin{equation}
\Delta=\mathbf{C}\setminus(-\infty,-1].
\end{equation}
The prefactor  exactly compensates the singularity of the Legendre function  and the product i regular there.
Points lying on the cut $[-\infty,-1]$ correspond to pairs of timelike separated real events on the de Sitter manifold $dS_d$.

The symmetry property of the Legendre functions of the first kind,
\begin{align}
P^\mu_{-\frac{1}{2}-i\kk}(z)
=
P^\mu_{-\frac{1}{2}+i\kk}(z),
\qquad
\kk\in\mathbb{C},\; z\in\Delta .
\end{align}
does not hold for the Legendre functions of the second kind. 
As a consequence, in the AdS case two distinct two-point functions can arise for a given mass, although they are not independent. 
Indeed, using \cite[Eq.~(3.3.1.4)]{Bateman}, one obtains the identity
\begin{align}
Q_{-\frac{1}{2}-\nu}^{\mu}(z)
=
Q_{-\frac{1}{2}+\nu}^{\mu}(z)
+
e^{i\pi\mu}\sin(\pi\nu)
\Gamma\!\left(\tfrac{1}{2}+\mu-\nu\right)
\Gamma\!\left(\tfrac{1}{2}+\mu+\nu\right)
P_{-\frac{1}{2}+\nu}^{-\mu}(z).
\end{align}
This relation implies that
\begin{equation}
W^{(d)}_{-\nu} (z)
=
W^{(d)}_{\nu} (z)
+
\frac{\sin(\pi\nu)\Gamma \left(\frac{d-1}{2}+\nu\right)
\Gamma \left(\frac{d-1}{2}-\nu\right)}{(2\pi)^{\frac d 2}}
(z^2-1)^{-\frac{d-2}{4}}
P_{-\frac{1}{2}+\nu}^{-\frac{d-2}{2}}(z) .
\label{pop}
\end{equation}

The second term on the right-hand side is \emph{regular on the AdS cut} $z\in[-1,1]$ and therefore does not contribute to the commutator. 
Thus the two solutions coincide at short distances. 
However, since this term grows increasingly rapidly as $|\nu|$ becomes large (see \cite[Eqs.~(3.9.2)]{Bateman}), the two solutions exhibit markedly different long-distance behavior.

The question of whether both solutions are physically admissible has been solved long ago: the answer is provided by the requirement of positive definiteness (unitarity) \cite{breit}. An alternative viewpoint invoke the possible self-adjoint extensions of a certain Schr\"odinger operator \cite{bertola1,bertola2}.
\vskip 10 pt

The Schwinger function is the restriction of the maximally analytic  two-point function  to non-coincident Euclidean points:
\begin{equation}
S(z_1^{\scriptscriptstyle{E}}, {z_2^{\scriptscriptstyle{E}}) = \left.W(z_1,z_2)\right|_{EAdS_d \times EAdS_d \setminus \delta}}
\end{equation}
where $\delta$ is the diagonal. The Euclidean propagator coincides with the two-point Schwinger function away from coincident points and is its distributional extension to the diagonal.
Choosing the points in Eq. (\ref{kgtp}) as follows 
\begin{align}
& z_1= e_d,
 \ \  z_2(u,\omega)= \left(
i \omega^{0} \sqrt{u^2-1},
\omega^1 \sqrt{u^2-1},
\ldots,
\omega^{d-1} \sqrt{u^2-1},
u
\right), \ \   u>1 , \ \ \omega^2 =1\label{uuu}
\end{align}
so that $ z_1\cdot 
z_2 (u,\omega) = u>1$, we  may write  the Euclidean propagator as follows
\begin{equation}
 S^{(d)}_{\lambda}(u) =  W^{(d)}_\lambda(z_1, 
z_2 (u,\omega)) =\frac
{e^{-i\pi\frac {d-2}2}}{(2\pi)^{\frac{d}2}} (u^2-1)^{-\frac
{d-2}4} Q^{\frac {d-2}2}_{\frac {d-2} 2+\lambda}(u). \label{kgtps}
\end{equation}

\section{Plane wave analysis of the two-point function: spacetime dimension  $d=2$}\label{sec9} 

With all ingredients in place, we begin by analyzing the plane-wave expansion of two-point functions within the two-dimensional spacetime $\wt {AdS_2}$. Though simpler than in the more general 
$d$-dimensional setting, this simpler case already illustrates many important features of plane-wave expansions while avoiding some of the  geometrical complications encountered in higher dimensions.

The fundamental object is  the two-point function 
\begin{equation}
\W^{(2)}_\l(z_1,z_2) =
c_2(\lambda)\int_{\gamma} ( z_1\cdot \zeta)^{\l} ( z_2\cdot \zeta)^{-\l-1}
d\mu_\gamma(\zeta), \ \ z_1\in {\wt\ZZ_-}, z_2\in \wt \ZZ_+.
\label{pw}\end{equation}
In this definition, the dependence on the integration cycle $\gamma$ is left implicit on the left-hand side. On the right-hand side, we have the integral of a closed differential form, where the cycle $\gamma$ is a curve on the complex cone $\mathbb{C}_2^{(c)}$ which generally depends on the points $z_1$ and $z_2$ that have been chosen. By construction, equation~(\ref{pw}) defines a solution of the Klein–Gordon equation with respect to both variables and is analytic in $\mathcal{Z}_- \times \mathcal{Z}_+$ (provided the integral converges).

Our objective is to determine those choices of the integration cycle $\gamma$ that yield an AdS-invariant result; in such cases, the resulting expression must necessarily be a linear combination of (\ref{kgtp}) and (\ref{kgtp2}).
The selection of the integration cycle  is crucial to enforce the AdS symmetry and obtain covariant quantizations of the Klein-Gordon field.

\subsection{A closed cycle and the special case of half-integer mass parameter}

Let us proceed by mimicking the de Sitter case \cite{bgm,bm} and consider the integral
(\ref{pw}) over a complete closed cycle $\gamma$ on the two-dimensional real cone $\CC_2$ that does not depend on the two points $z_1$ and $z_2$; let us consider the integral over  the upper component $\gamma^+_{S(2)}$ of the spherical  basis of $\CC_2$: 
\begin{eqnarray} &&\gamma_{S(2)}^{+} =\left \{\xi=(\sin\phi,1,\cos\phi), \ \ -\pi <\phi<\pi  \right\}, \ \ d\mu_{\gamma^+_{S(2)}}(\xi)= d\phi.\end{eqnarray} 
Properties 3 and  4 of Sect. \ref{cone} guarantee that  the integrand is always finite and therefore the integral converges. Unfortunately,  in general, the integrand is not univalued  on $\CC_2$; after one revolution it gets multiplied by the  global phase factor  $\exp(-4\pi i \lambda)$.
The conclusion is that  Stokes theorem cannot be applied to deform the contour and the result is not AdS invariant.

Exceptions occur when    $\lambda= l$ integer or $\lambda= l+\frac 12 $ half-integer. When $\lambda= l$  the integrals over the two adjacent half-circles exactly cancel each other, causing the total integral to vanish identically.
In contrast, when  $\lambda= l+\frac 1 2$ the integrand has a period of 
$\pi$ and the contributions from the two halves combine coherently to yield, in a nontrivial manner, the invariant maximally analytic solution (\ref{kgtp}):\begin{equation}
\W^{(2)}_{l+\frac 1 2}(z_1,z_2) = W^{(2)}_{l+\frac 12}(z_1,z_2)= \frac{1}{8\pi} \int_{\gamma_{S(2)}^{+}} ( z_1\cdot \xi)^{
l+\frac 12 
} ( z_2\cdot \xi)^{-l-\frac 32}
d\mu_{\gamma_{2+}}(\xi)= \frac 1{2\pi}  Q_{l+\frac 12}(z_1\cdot z_2). \label{8.18}
\end{equation} 

\subsection{General case: relative homology}
For general complex $\lambda$ integrating over a closed real cycle on the cone $\CC_2$ such as $\gamma^+_{S(2)}$  gives a non-invariant result.
 Extending the integration to cycles on the covering space of the cone does not resolve this issue and is physically unsound, as it would lead to an overcounting of possible momentum directions. In fact, even a complete cycle of the type considered above already entails such an overcounting, which partly accounts for the vanishing of the integral (\ref{pw}) when 
 $\lambda$  is an integer. 

\vskip 5 pt 
The solution for a general complex $\lambda$ is nevertheless given by Eq. (\ref{pw}), provided the integration cycle $\gamma$  is specified appropriately; here we generalize the results of \cite{bm2} where integer values of $\lambda$ were studied: for $\Re \lambda >-1$ the  cycle  $\gamma$ should  be open and have its endpoints on the plane orthogonal to $z_1$; more precisely, for $\Re \lambda >-1$ the cycle should belong to a  relative homology class of $H_{1}(\CC_-,\{\zeta:\,\zeta\cdot z_1 =0\})$;
for $\Re \lambda < 0$ the cycle should belong to a relative homology class of $H_{1}(\CC_+,\{\zeta:\,\zeta\cdot z_2 =0\})$; the two possible choices actually give the same results. 

\vskip 5 pt 
Let us examine the first option. For $\Re\lambda\geq0$ let us choose a relative cycle $\gamma(z_1)$ in $H_{1}(\CC_-,\{\zeta:\,\zeta\cdot z_1 =0\})$ i.e. a path in $\CC_-$ whose extremal  points belong to the plane $\zeta\cdot z_1 =0$:
    \begin{equation}
\W^{(2)}_\l(z_1,z_2) =
\frac{1}{4\pi}\int_{\gamma(z_1)} ( z_1\cdot \zeta)^{\l} ( \zeta\cdot z_2 )^{-\l-1}
d\mu_{\gamma(z_1)}(\zeta), \ \ z_1\in {\ZZ_-}, z_2\in \ZZ_+
\label{pw2}.\end{equation}
    
  \begin{itemize}
      \item
The value of the integral is independent of the chosen representative within the relative homology class. This follows from the fact that the integrand is a locally well-defined closed differential form, and that it vanishes on the relative boundary.
   \item Convergence of the integral then allows to relax the condition to  $\Re\lambda\geq -1$. The independence on the chosen representative in the relative homology class remains true.

   \item
AdS invariance follows by construction from the relation  $\gamma(g z_1)=g\gamma(z_1)\subset \CC_- $:
\begin{eqnarray}
\W^{(2)}_\l(g z_1,g z_2) =
\frac{1}{4\pi}\int_{g \gamma(z_1) } ( z_1\cdot g^{-1} \zeta)^{\l} (g^{-1}\zeta\cdot  z_2)^{-\l-1}
d\mu_{g\gamma(z_1)}(\zeta)= \cr= \frac{1}{4\pi}\int_{\gamma(z_1) } ( z_1\cdot \zeta)^{\l} (\zeta\cdot  z_2)^{-\l-1}
d\mu_{\gamma(z_1)}(\zeta).
\end{eqnarray} \end{itemize}
$\W_\l(z_1,z_2)$ is thus a  function  of one complex variable, namely of the scalar product  $\scalar{z_1} {z_2}\in \Delta$;  it is univalued only for integer $\l=l$. o evaluate this function, we choose the two points in the respective tuboids  as follows:
\begin{eqnarray}
z_1=(\sin(-i v),0,\cos
(-i v)) , \ \ v\geq 0, 
&& z_2 =(\sin(i u),0,\cos
(i u)), \ \ u>0. \label{config}
\end{eqnarray}
A representative of the relevant relative  homology class 
is the following cycle in $\CC_-$:
\begin{eqnarray}
\gamma(z_1) =\left \{(\sin(\phi-iv),1,\cos
(\phi-iv)), \ \ -\pi/2 <\phi<\pi/2  \right\} \label{cycle0}
\end{eqnarray}
In the special case when $v=0$ the first point $z_1$ is  on the real manifold and coincides with the base point $e_2$; the corresponding relative cycle is a subset of $\gamma^+_{S(2)}$ (see Figure (\ref{fig:1})) 
\begin{eqnarray}
\gamma^+_{S(2)}(e_2) =\left \{(\sin\phi,1,\cos
\phi), \ \ -\pi/2 <\phi<\pi/2  \right\}=  \left \{\xi \in \gamma^+_{S(2)}: e_2\cdot \xi = \cos \phi\geq 0  \right\}. \label{cycle00}
\end{eqnarray}
In all  cases Eq. (\ref{pw2}) becomes 
\begin{eqnarray}
\W^{(2)}_\l(z_1,z_2)= \frac{1}{4 \pi}\int_{-\frac \pi 2 }^{\frac \pi 2} ( \cos\phi)^{\l}  ( \cos(\phi -i (u+v)))^{-\l-1}
d\phi  \label{int0}
\end{eqnarray}
and it is clear that we can set $v=0$ without loss of generality. The change of variables $\phi= \arctan (\sinh t)$ gives
\begin{equation}
\W^{(2)}_\l(z_1,z_2)= 
 \frac{1}{4 \pi} \int_{-\infty }^{\infty } {(\cosh u+i\ \sinh t  \ \sinh u)^{-\l-1}} dt  = \frac{1}{2\pi}Q_\l(\cosh u) 
\end{equation}
i.e.
\begin{equation}
\W^{(2)}_\l(z_1,z_2)= 
W^{(2)}_\lambda(z_1,z_2)  = 
\frac{1}{2\pi} Q_\l(\scalar{z_1}{z_2}). \label{ident}
\end{equation}
A similar formula holds true for the permuted two-point function $W_\l(x_2,x_1)$, 
but the primitive domain of analyticity is the opposite tuboid ${\ZZ_+}\times {\ZZ_-}$. 
Taking the difference of the boundary values gives the commutator:
\begin{equation}
C^{(2)}_\l(x_1,x_2) = W^{(2)}_\l(x_1,x_2) - W^{(2)}_\l(x_2,x_1).
\end{equation}
 For a closer look into Eq. (\ref{pw2})  we may  change the integration variable in (\ref{int0}) by taking $z= i e^{i\phi}$: 
\begin{eqnarray}
W^{(2)}_\lambda(\cosh u)
=\frac{ e^{(\lambda+1) u}}{2\pi}\int_{\gamma^+} \left(\zu^2-1\right)^\lambda \left(e^{2 u}\zu^2-1\right)^{-\lambda-1}d\zu \label{rr}.
\end{eqnarray}
 The integration cycle (\ref{cycle0}) is mapped  counterclockwise onto the upper unit semicircle in the complex  $z$-plane: $\gamma^+=\{|z|=1,\  \Im z>0\}$. For  $\zu=x+i y$ the  null vector 
\begin{eqnarray}
\zeta(z) =\left(-\frac{\zu^2+1}{2 \zu},1,-\frac{ i(\zu^2-1)}{2 \zu}\right)
\label{parm}
\end{eqnarray} 
is such that 
\begin{eqnarray}
  && (\Im \zeta)^2 =\frac{\left(x^2+y^2-1\right)^2}{4 \left(x^2+y^2\right)}\geq 0, \ \ \  \epsilon( \zeta) =\frac{\left(1-x^2-y^2\right) \left(1+x^2+y^2\right)}{4 \left(x^2+y^2\right)}.
\end{eqnarray}
This implies that  $\zeta(z)$ belongs to $\CC_+$ when $|\zu|<1$ and to $\CC_-$ when $|\zu|>1$. 
This shows that $\zeta(z)$ lies in $\CC_+$ when $|z| < 1$, and in $\CC_-$ when $|z| > 1$.
A deformation of the contour is allowed as long as the integration path remains within the closure of $\CC_-$.
Since the integral over the upper semicircle
\[
\gamma^+(r) = \{\, |z| = r,\, \Im z > 0 \,\}
\]
vanishes as $r \to \infty$, we can replace $\gamma^+$ in Eq.~(\ref{rr}) by the two half-lines
\[
\{\, |x| > 1,\, y = 0 \,\},
\]
and thus obtain
\begin{eqnarray}
W^{(2)}_\lambda(\cosh u)
&=&\frac{e^{(\lambda+1) u}}{\pi}\int_{1}^\infty \left(x^2-1\right)^\lambda \left(e^{2 u} x^2-1\right)^{-\lambda-1}dx= \cr &=& \frac{ \Gamma(\frac 12)  \Gamma (\l+1) \,
   }{2 \pi\Gamma
   \left(\l+\frac{3}{2}\right)} e^{-(\l+1) u}{}_2F_1\left(\frac{1}{2},\l+1;\l+\frac{3}{2};e^{-2 u}\right). \label{rr1}
\end{eqnarray}
Comparing this alternative  
representation  of the two-point function 
with Eq. (\ref{ident}) we deduce a special case of a known (see \cite[Eq. 45, page  136]{Bateman}) and useful identity which we reproduce here for the reader's convenience: 
\begin{equation}
Q^\mu_\nu (\cosh u) = \frac{ 2^{\mu } e^{i \pi  \mu }\sqrt{\pi }  \Gamma (\mu +\nu +1) e^{-u (\mu +\nu +1) } (\sinh
   u)^{\mu } \, _2F_1\left(\mu +\frac{1}{2},\mu +\nu +1;\nu +\frac{3}{2};e^{-2
   u}\right)}{\Gamma \left(\nu +\frac{3}{2}\right)}.\label{Qalphabeta}
\end{equation}

An integral representation with integration over a compact interval can also be obtained via the conformal transformation $\hat{z}=1/z$,  which exchanges  the inner and the outer regions of the unit circle Under this mapping,    $\zeta(\hat{z})$ belongs to $\CC_+$ when $|\hat{z}|>1$ and to $\CC_-$ when $|\hat{z}|<1$. 

On the unit circle  we have $\hat{z}= -i e^{-i \phi}$ so that   the integration cycle is mapped counterclockwise onto the lower half $\gamma^-$ of the unit circle $|\hat{z}|=1$. By contour distortion taking place within $\CC_-$  we may  replace $\gamma^-$  with the compact interval $[-1,1]$ thus obtaining 
\begin{equation}
{\W^{(2)}_\l}(\cosh u)
=\frac{ e^{-(\lambda+1) u}}{2\pi}\int_{-1}^1 \left(1-x^2\right)^\lambda \left(1-e^{-2 u} x^2\right)^{-\lambda-1}dx. 
\label{rr22}
\end{equation}

\vskip 5 pt

For $\Re \lambda<0$ we may also consider relative cycles in $\CC_+$  having their endpoints on the plane $\xi \cdot z_2=0$: 
\begin{equation}
{\W^{(2)}_\l}(z_1,z_2) = \frac{1}{4\pi}
\int_{\gamma(z_2)} ( z_1\cdot \xi)^{\l} (\xi\cdot z_2)^{-\l-1}
d\mu_{\gamma(z_2)}(\xi)
\label{p.12l}\end{equation}
A representative of the relevant  homology class is the complex cycle
\begin{eqnarray}
\gamma(z_2) =\left \{\xi 
 =(\sin(\phi+i u),1,\cos
(\phi+i u)), \ \ -\pi/2 <\phi<\pi/2  \right\}. \label{cycleplus}
\end{eqnarray}
 The change of variable $z= i e^{i\phi}$ plus contour distortion give
\begin{eqnarray}
{\W^{\sharp(2)}_\l}(\cosh u) 
= \frac{e^{\lambda u}}{4\pi} \int_{-1}^{1} \left(1-e^{-2u}x^2\right)^\lambda \left(1- x^2\right)^{-\lambda-1}dx  \label{laltro}
\end{eqnarray}
By comparing the latter expression with Eq. (\ref{rr22})  we get 
\begin{eqnarray}
&& {\W^{\sharp(2)}_\l}(z_1,z_2)  =  \frac{1}{2\pi} Q_{-\lambda-1} (z_1,z_2) \label{laltro2}
\end{eqnarray}
which coincides with the previous one in the allowed range of values of $\lambda$.

\subsection{Application to Legendre functions: a multiplication theorem  for $Q_\lambda$}

The key new feature of Eq. (\ref{pw2}) is that it provides an integral representation of 
$Q_\lambda(z_1\cdot z_2)$ 	
  in terms of two distinct points through their scalar product, opening up a wide range of potential applications. Equation (\ref{pw2}) can be used to derive numerous new integral representations of Legendre functions with more complicated arguments (see, for instance, Eqs. (\ref{1010a}) and (\ref{28})). Moreover, Eq. (\ref{pw2}) can also be employed to generate new formulas for Legendre functions of the second kind.

  To give an example  let us  choose  $z_1=e_d$ and  $z_2 =(x \sin(i u),\sqrt{x^2-1},x\cos
(i u))$;  Eq. (\ref{pw2}) gives 
\begin{eqnarray}
Q_\l(x\, \cosh u)= \frac{1}{2}\int_{-\frac \pi 2 }^{\frac \pi 2} ( \cos\phi)^{\l}  \left( x \cos(\phi -i u)-\sqrt{x^2-1}\right)^{-\l-1}
d\phi.  \label{int01b}
\end{eqnarray}
By replacing the second factor in the integrand on the right-hand side with its binomial series, we are led to consider a series of integrals. These integrals can be evaluated using yet another new integral representation for associated Legendre functions of the second kind, which is derived in Eq. (\ref{oopp}) below:
\begin{eqnarray}
\int_{-\frac \pi 2 }^{\frac \pi 2} ( \cos \phi )^{\l}  (  \cos(\phi -i u))^{-\l-1-n}
d\phi= \frac{2^{1+\frac{n}{2}} \sqrt{\pi } \Gamma (1+\lambda )  e^{-\frac{1}{2} i n \pi }
   Q_{\frac{n}{2}+\lambda }^{\frac{n}{2}}(\cosh (u))}{\Gamma \left(\frac{1}{2}-\frac{n}{2}\right) \Gamma (1+n+\lambda ) (\sinh u)^{\frac{n}{2}}}.  \label{int01bis}
\end{eqnarray}
What follows is a nontrivial — and possibly previously unknown — multiplication theorem for associated Legendre functions of the second kind:
\begin{eqnarray}
Q_\l(x\, \cosh u)=  \frac{1}{ x ^{1+\lambda }}\sum_{n=0}^\infty\frac{1}{ \Gamma (1+ n)  }    \, \left(1-\frac{1}{x^2} \right)^{ n}  
  \frac{Q_{n+\lambda }^n(\cosh u)}{  (2 \, \sinh u)^{n} } \label{int01}, \ \ x>1.
\label{examp} \end{eqnarray}

 \section{Bulk-to-boundary limit and the shadow transformation}
 \label{sec10}
 The plane-wave expansion  (\ref{pw2}) offers the possibility to   study the boundary behaviour of the two-point function in a very simple way  by  sending  $z_1$ to the boundary of $\wt {AdS_2}$ here identified with the covering of real cone $\CC_2$. To this aim let us choose
\begin{eqnarray}  z_1(t,\kkappa)  =\frac{\hat z_1(t,\kkappa)}{ \kkappa}  
=\frac 1 \kkappa {\left(\sqrt{ \kkappa^{2}+1} \sin t,1,\sqrt{ \kkappa^{2}+1}  \cos t\right)}  , \ \ z_2 =(\sin(i u ),0,\cos(i u)) \\
\gamma(z_1) =\{
\xi= (\sin\phi,1,\cos\phi), \ \ t-t_\kkappa<\phi  <t+ t_\kkappa\},\ \ t_\kkappa = \arccos\left((\kkappa^2+1)^{-\frac 12}\right).\end{eqnarray}
Note that   $z_1(t,\kkappa)\cdot z_1(t,\kkappa) = 1$  while   $\hat z_1(t,\kkappa)\cdot \hat z_1(t,\kkappa)= \kkappa^2$ and we are going to take the limit $\kkappa\to 0$. 
Suppose that $\kkappa>0$; when $\kkappa\to 0$ the integration is over the small interval $[t-\kkappa,t+\kkappa]$; the plane-wave representation (\ref{pw2}) of the two-point function  immediately gives
\begin{eqnarray}
&&
W^{(2)}_\l(z_1(t,\kkappa),z_2)
\simeq\frac{(\cos(t-i u))^{-\l-1}}{4\pi \kkappa^{\lambda}}\int_{-t_\kkappa}^{t_\kkappa} \left( \sqrt{\kkappa^2+1} \cos \phi -1\right)^{\l} d\phi  \cr&&  \simeq 
\frac{ 2^{-\lambda-2 } \Gamma (\lambda +1) \kkappa ^{ \lambda +1}}{\sqrt{\pi }\Gamma
   \left(\lambda +\frac{3}{2}\right)} (\hat z(t,\kkappa)\cdot z_2)^{-\l-1}. \label{ppoo}
\end{eqnarray}

For $\kkappa<0$ the integration path tends instead to the whole circle $[t-\pi,t+\pi]$ but the result does not change. 
For $\lambda>-1$ the rhs of (\ref{ppoo}) tends to zero but the following expression has a limit
\begin{eqnarray}
\lim_{\kkappa \to 0^\pm} \kkappa^{-\lambda-1}W^{(2)}_\lambda(z_1(t,\kkappa),z_2)
 =  \frac{2^{-\lambda-2} \Gamma(\lambda+1)}{\sqrt{\pi}\Gamma(\lambda+\frac 32)}\left({\scalar{\xi_\pm }{z}}\right)^{-\lambda-1} 
\end{eqnarray}
where 
\begin{eqnarray}
\xi_\pm=\pm (\sin t, 1, \cos t)
\end{eqnarray}
are the null vectors of $\CC_2$ representing the relevant limit points of the boundary.
 The bulk-to-boundary two-point function is thus proportional to a plane wave.

\subsection{The shadow transformation}
\vskip 5 pt

If we now let the second point $z_2$	
  approach the boundary while keeping 
$z_1$	
  fixed, we obtain an integral formula relating the two waves conjugated by the involution (\ref{involution}). Such formulas are sometimes referred to in the literature as “shadow” transformations.
Let us therefore choose  
\begin{eqnarray}
z_1(v)=(\sin(-i v),0,\cos
(-i v)) , \ \ v>0,  
\ \  \ \ z_2(t,\kkappa) = \frac{\hat z_2(t+i s,\kkappa)}{ \kkappa}, \ \ s>0  \label{config34}
\end{eqnarray}
and  the cycle  (\ref{cycle0}) in $\CC_-$:
\begin{eqnarray}
W^{(2)}_\l(z_1(v),z_2(t,\kkappa))&=& \frac{{\kkappa }^{\l+1}}{4 \pi}\int_{-\frac \pi 2 }^{\frac \pi 2}( \cos \phi)^{\l}   \left( {\sqrt{\kkappa ^2+1} \cos (t+i(s+ v)-\phi )-1}\right)^{-\l-1}d\phi  \label{int000}  \cr &\simeq& \frac{ \kkappa ^{\lambda +1} \Gamma (\lambda +1) (\cos (t+i(s+ v)))^{-\lambda -1}}{2^{\lambda +2}\sqrt{\pi
   } \Gamma \left(\lambda +\frac{3}{2}\right)}.
\end{eqnarray}
By using AdS invariance and letting $\kkappa\to 0$ we deduce that for  $ z\in  \ZZ_- $ and $\zeta\in {\CC_+}$ there holds the following identity
\begin{equation}
(z\cdot \zeta)^{-\l-1} =
C_2(\lambda)
\int_{\gamma(z)} ( z\cdot \zeta')^{\l} (\zeta'\cdot \zeta)^{-\l-1}
d\mu_\gamma(\zeta')
\label{p.12aa}\end{equation}
valid for $\Re \lambda>-1$, where  $\gamma(z)$ is a cycle in $\CC_-$  relative to the plane $\zeta' \cdot z=0$  and \begin{equation} C_2(\l) =  \frac{2^{\l} \Gamma \left(\l+\frac{3}{2}\right)}{\Gamma \left(\frac{1}{2}\right) \Gamma (\l+1)}.\end{equation}
This formula provides a concrete realization of the “shadow” transformation  establishing an explicit integral relation between waves conjugated by the involution (\ref{involution}). A completely analogous formula holds for $z\in \ZZ_+$ and $\zeta\in {\CC_-}$.
\vskip 5 pt

Putting everything together we finally arrive at the following more symmetric integral representation of the two point function: let $z_1\in \ZZ_-$, $z_2\in \ZZ_+$; $\gamma(z_1)$ a cycle in $\CC_-$ relative to the plane $\scalar{z_1}{\zeta_1}=0$;  $\gamma(z_2)$ a cycle in $\CC_+$  relative to the plane $\scalar{z_2}{\zeta_2}=0$; then, for $\Re \lambda>-1$ there holds the following plane wave expansion of the two-point function:
  \begin{equation}
W^{(2)}_\l(z_1,z_2) =
\frac{C_2(\l)} {4\pi}\int_{\gamma_1(z_1)} \int_{\gamma_2(z_2)} ( z_1\cdot \zeta_1)^{\l} (\zeta_1\cdot \zeta _2)^{-\l-1}
 (  \zeta_2\cdot z_2)^{\l} d\mu_{\gamma(z_1)}(\zeta_1)d\mu_{\gamma(z_2)}(\zeta_2). 
\end{equation}

\section{General spacetime dimension $d$} \label{generald}
\label{sec11}

As is natural in the case of a Fourier-like plane-wave analysis, the results of the previous paragraph should be generalizable  to arbitrary dimension $d$. We expect that the Fourier-like formula (\ref{pw2}) should look the same in every dimension $d$; this means that, as anticipated in the Introduction, for $z_1\in\wt \ZZ_-$ and $z_2\in\wt \ZZ_+$ it should 
\begin{align}
    W^{(d)}_\lambda(z_1,z_2)=
    c_d(\lambda) \int_{\gamma(z_1)} (z_1\cdot \zeta)^\lambda ( z_2\cdot \zeta)^{-\lambda-d+1} d\mu_\gamma(\zeta) \label{pwdu}
\end{align}
where $\gamma(z_1)$ in $H_{d-1}(\CC_-,\{\zeta:\,\zeta\cdot z_1 =0\})$ is a cycle relative to the hyperplane $\zeta\cdot z_1 =0$, 
the  measure is the restriction to the chosen  cycle  
of the invariant volume form on the cone  and the constant $c_{d}(\lambda)$ is chosen by  canonical normalization  to reproduce   Eq. \eqref{kgtp}.

Since the two-point function  Eq. (\ref{pwdu}) is AdS invariant, we may compute it for the particular configuration
\begin{align}
    z_1&=(\sin(-iv),\vec 0,\cos(-iv)), \quad v>0,\quad z_1\in \mathcal Z_-, \\
    z_2&=(\sin(iu),\vec 0,\cos(iu)), \quad \quad\ \,  u>0,\quad z_2\in \mathcal Z_+; \label{pop12}
\end{align}
the integration cycle in the relevant homology class is chosen as follows:
\begin{align}
    \gamma(z_1)=\left\{\zeta=(\sin(\phi-iv),\vec n,\cos(\phi-iv)), \ \  -\frac\pi2<\phi<\frac \pi2,\ \  \vec n \in S_{d-2}\right\} \label{cyc2}.
\end{align}
In the special case where $v=0$, i.e. when $z_1=e_d$, the integration cycle can be taken real; in particular  one can chose a relative cycle contained in  $\gamma_S$ which,  at variance with the two-dimensional cycle $\gamma_{S(2)}$, is  a connected subset of $\CC_d$:
\begin{eqnarray}
\gamma_S(e_d) =\left\{\xi \in \gamma_{S}: e_d\cdot \xi = \cos \phi\geq 0  \right \}. \label{cycle00d}
\end{eqnarray}
In the above configuration the integration over $S_{d-2}$ is trivial and  we get
 \begin{align}
    W^{(d)}_\lambda(z_1,z_2)
    =& \frac {2\pi^{\frac {d-1}2}c_{d}(\lambda) }{\Gamma(\frac {d-1}2)} \int_{-\frac \pi2}^{\frac \pi2} (\cos\phi)^\lambda (\cos(\phi-iu))^{-\lambda-d+1} d\phi  \label{pppp} \\  = &-\frac {2^d i \pi^{\frac {d-1}2}c_{d}(\lambda) }{\Gamma(\frac {d-1}2)}  e^{u(\lambda+d-1)}\int_{\gamma^+} \left(-\frac{i \left(z^2-1\right)}{z}\right)^{\lambda } \left(-\frac{i \left(e^{2 u}
   z^2-1\right)}{z}\right)^{-d-\lambda +1} \frac{dz}{z};
 \end{align}
in the second step we changed the integration variable to $z=i e^{i\phi}$; the  integral is over the upper unit semicircle $\gamma^+$ in the complex  $z$-plane counterclockwise. By contour deformation  $\gamma^+$ may be replaced  by the two  half-lines $\{z= x+iy,\ |x|>1, y=0\}$. There is however a  difference w.r.t. the two-dimensional case:  the two integrals  here differ by a phase factor; by  adding their contributions we get   
\begin{eqnarray}
 &&  \int_{-\frac \pi2}^{\frac \pi2} (\cos\phi)^\lambda (\cos(\phi-i u ))^{-\lambda-d+1} d\phi =  \cr && = \frac{   \pi\Gamma (\lambda +1) e^{-u (d+\lambda -1)} \, _2F_1\left(\frac{d-1}{2},d+\lambda -1;\frac{d+1}{2}+\lambda ;e^{-2 u}\right)}{2^{1-d}\Gamma  \left(\frac{3}{2}-\frac{d}{2}\right) \Gamma \left(\frac{d+1}{2}+\lambda \right)} =  
  \cr &&
   =\frac{2^{\frac d 2} \sqrt{\pi } \Gamma (1+\lambda )
  }{\Gamma \left(\frac{3}{2}-\frac{d}{2}\right) \Gamma (d-1+\lambda
   )}  \,  (\sinh
  u) ^{\frac{2-d}{2}}\,  e^{-\frac{i  \pi(d-2)}{2}  } Q_{\frac{d-2}{2}+\lambda }^{\frac{d-2}{2}}(\cosh u)    \label{oopp}
\end{eqnarray}
where, in the last step, we used Eq. \eqref{Qalphabeta}.\footnote{The above calculation provides another yet unknown (at least to me) integral representation of the Legendre function of the second kind that it is  worth singling out:
\begin{eqnarray}
   e^{-i\pi\mu} Q^\mu_\nu(\cosh u) = \frac{2^{-\mu -1} \Gamma \left(\frac{1}{2}-\mu \right) \Gamma (\mu +\nu +1)  }{\sqrt{\pi } \Gamma
   (-\mu +\nu +1)} (\sinh u)^{\mu
   } \int_{-\frac \pi 2}^{\frac \pi 2} (\cos \phi )^{ -\mu +\nu} (\cos ( \phi-i u  ))^{-\mu -\nu -1} d\phi
\end{eqnarray}}
By comparing Eqs. 
 \eqref{kgtp}, \eqref{pppp} and  \eqref{oopp} we finally get the value of the normalization constant 
\begin{equation}
    c_{d}(\lambda)=-\frac{  \pi ^{1-d} \Gamma
   (\lambda+d -1)}{ 2^{d+1}\cos \left(\frac{\pi  d}{2}\right) \Gamma (\lambda +1)}.
   \end{equation}
The above formula gives $c_2(\lambda)= 1/{(8\pi)}$ whereas in Eq. \eqref{pw2} we had $ 1/({4\pi})$. Again, this is because in the two-dimensional case $\gamma_2$ has two disconnected components and  we took the integral just  on a cycle contained in one  component $\gamma_{2+}$; the integral over the second component would produce an identical contribution.  This  peculiarity of the two-dimensional case  disappears in the general case (see also Sect. \ref{poincare} for another important choice of integration cycle). 

When $d=2n+1$ is an odd integer the constant $c_d(\lambda)$ diverges but, at the same time,  the integral vanishes; the  corresponding two-point function may be  obtained as a limit:
\begin{eqnarray}
 W^{(2n+1)}_\lambda(z_1,z_2)=
\frac{(-1)^{n+1} \Gamma(\lambda+2n
   )}{(2 \pi )^{d} \Gamma (\lambda +1)}
\int_{\gamma(z_1)} (z_1\cdot \xi)^\lambda (\xi\cdot z_2)^{-\lambda-2n}\log(\xi\cdot z_2) d\mu_\gamma(\xi). \label{pwduodd} \
\end{eqnarray}
\vskip 10 pt 
As before we may use Eq. (\ref{pwdu}) to generate non trivial identities among Legendre functions.  For example we may generalize the previous multiplication theorem (\ref{examp}) by  choosing  $z_1=e_d$ and 
\begin{align}
    z_2&=\left(x \sin(iu),\sqrt{x^2-1}\  \vec {\rm n} ,x\cos(iu)\right), \quad u>0, \ \ |\vec n|=1.\label{pop3}
\end{align}
Proceeding as for Eq. (\ref{int01}) we get 
\begin{align}
& Q_{\nu}^{\mu}(x\,  \cosh u) 
\frac{(x^2 \, \cosh^2 u-1)^{\frac\mu 2} }{x ^{1+\mu+\nu}}\sum_{n=0}^\infty  \frac{\left(\frac 1 2-\frac{1}{2\, x^2} \right)^{ n}  }{ \Gamma
   (1+n) (\sinh u)^{\mu+n}} Q_{\nu+n }^{\mu+n}(\cosh (u))
 \end{align}

 \subsection{The shadow transformation}
 
If in Eq. (\ref{pwdu}) we let $z_2$  move towards the boundary by keeping $z_1$ fixed,   we discover the general "shadow" transformation relating  the two waves conjugated  by the involution (\ref{involution}). To this aim we choose 
\begin{eqnarray}
&&\hat z_2(t+iu,\vec n', \kkappa)=\left(\sqrt{ \kkappa^2+1} \sin (t+i u),\vec n',\sqrt{ \kkappa^2+1}  \cos (t+iu)\right) \in \AdS(R) \cr && z_2(t+i u ,\kkappa) = \frac{\hat z_2(t+i u,\kkappa)}{ \kkappa}, \ \ u>0  \label{config3}
\end{eqnarray}
and  the cycle  (\ref{cyc2}) in $\CC_-$; we get
\begin{eqnarray}
 &&   W^{(d)}_\l(z_1(v),z_2(t+iu ,\kkappa))=\cr 
 &&    = 
 c_{d}(\lambda) \kkappa ^{\l+d-1} \int_{-\frac \pi2}^{\frac \pi2} \int_{S^{d-2}} (\cos\phi)^\lambda  (\cos (t+i(u+ v)-\phi )-\vec n\cdot \vec n')^{-\lambda-d+1} d\phi d\vec n\cr &&  = 
 \frac{ \kkappa ^{\l+d-1}(z_1\cdot \hat z_2)^{1-d-\lambda } \Gamma (d-1+\lambda ) \, }{2^{d+\lambda } \pi ^{\frac{d-1}{2}} \Gamma \left(\frac{d+1}{2}+\lambda
   \right)}
\end{eqnarray}
Letting $\kkappa\to 0$ 
and taking into account the AdS invariance of the above expression we deduce that
\begin{eqnarray}
   {(z\cdot \zeta')^{1-d-\lambda }  } =        C_d(\lambda)\int_{\gamma(\zeta)} ( z\cdot \zeta)^{\l} (\zeta\cdot \zeta')^{-\l-d+1}
d\mu_\gamma(\zeta)
\end{eqnarray}
valid for $z\in \wt \ZZ_-$, $\zeta'\in \ovl \CC_+$ and $\gamma(\zeta)\in H^1(\CC_-,\{\zeta:\,  z\cdot \zeta=0\})$. A similar formula holds for $z\in \ZZ_+$ and $\zeta'\in {\CC_-}$. In both cases  the normalization constant takes the following value:
\begin{equation}
C_d(\lambda) = - \frac{2^{\lambda-1 }  \Gamma \left(\frac{d+1}{2}+\lambda
   \right)} { \pi ^{\frac{d-1}{2}}\cos \left(\frac{\pi  d}{2}\right) \Gamma (\lambda +1)}
\label{p.126}\end{equation}
 Note that again we have an extra 1/2 factor when we set $d=2$ w.r.t. Eq. (\ref{p.12aa}) because here we are integrating on the two components over the disconnected sphere $S_0$.

Putting everything together we finally arrive at the following integral representation of the two point function: let $z_1\in \ZZ_-$, $z_2\in \ZZ_+$; $\gamma(z_1)$ a cycle in $\CC_-$ relative to the plane $\scalar{z_1}{\zeta_1}=0$;  $\gamma(z_2)$ a cycle in $\CC_+$  relative to the plane $\scalar{z_2}{\zeta_2}=0$; then, for $\Re \lambda>-1$ there hold the following integral representation of the two-point function:
  \begin{equation}
W^{(d)}_\l(z_1,z_2) =
{c_d(\l)}{C_d(\l)} \int_{\gamma_1(z_1)} \int_{\gamma_2(z_2)} ( z_1\cdot \zeta_1)^{\l} (\zeta_1\cdot \zeta _2)^{-\l-d+1}
 (  \zeta_2\cdot z_2)^{\l} d\mu_{\gamma(z_1)}(\zeta_1)d\mu_{\gamma(z_2)}(\zeta_2). 
\end{equation}

\def\K{\varkappa}
\def\E{ E}
\def\q{\omega}
\def\P{ P}
\def\y{ {\rm y}}
\def\Sp{\hbox{Spec}(\square_\Y)}

\newcommand{\be}{\begin{equation}}
\newcommand{\ee}{\end{equation}}
\newcommand{\ba}{\begin{eqnarray}}
\newcommand{\ea}{\end{eqnarray}}
\newcommand{\pa}{\partial}
\newcommand{\llambda}{\rho}
\renewcommand{\Re}{\mathrm {Re} \,}
\renewcommand{\Im}{\mathrm {Im} \,}

\def\r{t}
\def\bra#1{\le\langle{#1}|}
\def\vrul{\rule[20pt]{0pt}{0pt}}
\def\ket#1{{#1}\ri\rangle}
\def\bra#1{\le\langle{#1},}
\def\bea{\begin{eqnarray}}
\def\eea{\end{eqnarray}}
\def\le{\left}
\def\ri{\right}
\def\l{\rho}

\def\bes{$$}
\def\ees{$$}
\def\tensor{\otimes}
\def\o{\omega}

\def\eeta{\chi}
\def\M{AdS}

\section{The Poincar\'e patch $AdS_M$}
\label{sec12}
In this and the subsequent sections, we decompose the plane waves (\ref{waves}) and diagonalize the two-point function (\ref{pwduint}) using the Poincaré coordinate system. This framework opens up new opportunities to compute Feynman diagrams directly in Lorentzian AdS space, rather than relying on the conventional Euclidean approach.

We denote by $AdS_M$ (with $M$ standing for Minkowski) the region of the $d$-dimensional manifold ${\AdS}$ - or equivalently, of its universal covering space $\widetilde{\AdS}$ - that is foliated by the family of hyperplanes
\begin{eqnarray}
    \Pi_u= \{ x^{d-1}+ x^d =  1/u= e^v>0\}.
\end{eqnarray}
The following  system parametrizes the leaves by Lorentzian (Poincar\'e) inertial coordinates
\begin{equation}
x(\x,u)= \left\{\begin{tabular}{lclcll}
 $x^{\mu} $ & =& $\frac{ 1}{u}\x^\mu  $ &=& $e^{ {v}} \x^\mu, \ \ \ \ \mu=0,1,...,d-2 $\cr
 $x^{d-1} $  & =& $\frac{1-u^2}{2u} + \frac {1}{2 u} \x^2$  &=& $\sinh  {v} + \frac 12 e^{ {v}} \x^2 $ \cr
 $x^{d}$&  =& $\frac{1+u^2}{2u} -
\frac {1}{2 u} \x^2$&=&$ \cosh  {v} - \frac 12 e^{ {v}} \x^2$ 
\label{coordinates}
\end{tabular}\right.
 \end{equation}
where 
\begin{equation}
\x^2=\x^2_M = \x\cdot \x=\eta_{\mu\nu}\x^\mu\x^\nu,  \ \  \ \ d s_M^2 ={(d\x^0)}^2- {(d\x^1)}^2- \cdots -{(d\x^{d-2})}^2
\end{equation}
are the scalar product and the interval in a $(d-1)$-dimensional Minkowski space $M_{d-1}$. 
In Poincar\'e coordinates, the AdS scalar product and interval are expressed as follows:
\begin{eqnarray}
&& (x-x')_{AdS}^2 = \frac{\left(\x-\x'\right)_M^2 - ({u^2-{u'})^2}  }{uu'} , \ \ 
\label{7}\\
&& {\mathrm d}s^2 = \left.  {\left(dx^{0}\right)}^2 -{\left(dx^{1}\right)}^2-\ldots+{\left(dx^{d}\right)}^2\right|_{AdS} 
= \frac{1}{u^2}({{\mathrm d}s^2_{{M}}-{\mathrm d} u^2}). \label{metric1}
\end{eqnarray}
 Eq. (\ref{7}) implies that
\begin{equation}
(x(u,\x)-x(u,\x'))_{AdS}^2 = \frac 1 {u^2}(\x-\x')_M^2.
\label{lll}
\end{equation}
This equation shows that the Lorentzian geometry of the leaves is consistent with that of the ambient AdS space; that is, the causal ordering within any hyperplane $\Pi_u$ can be understood either in the Minkowskian sense of $\Pi_u$ or in the context of the ambient AdS spacetime. The manifold $AdS_M$ is strongly causal, though this comes at the cost of geodesic incompleteness.

The second crucial property  is that the the complex leaves $\Pi_u^{(c)}$ contain copies of the Minkowsian tubes 
\begin{equation}
    T_\pm = \{\z=\x\pm i \y \in M_{d-1}^{(c)}: \y^2>0 , \ \y^0  \gtrless 0 \}
\end{equation}
which are (respectively) contained in the ambient AdS tubes $\ZZ_\pm$.

 This can be stated as follows: consider a partial complexification of $AdS_M$ in which the Poincar\'e coordinates $\mathbf{x}^\mu$ in Eq.~(\ref{coordinates}) are extended to complex values, while the coordinate $u$ remains real and positive:
\begin{equation}
z(\z,u)=\left\{\begin{tabular}{lcl}
 $z^{\mu} $ & =& $\frac{ 1}{u}\z^\mu  $\cr
 $z^{d-1} $& =& $\frac{1-u^2}{2u} +
\frac {1}{2 u} \z^2$ \  ,  \cr
 $z^{d}$&= & $\frac{1+u^2}{2u} -
\frac {1}{2 u} \z^2$
\label{coordinatesc}
\end{tabular}\right. \ \   \z^2= \eta_{\mu\nu}\z^\mu\z^\nu, \ \ \ u>0.
\end{equation}
 The leaves at fixed $u>0$ are copies of $M_{d-1}^{(c)}$.  Suppose now that $\z= \x+i\y$ and that $\y^2_M>0$, i.e. $\z$ belongs to either $T^+$ or $T^-$; it follows immediately that   
\begin{eqnarray}
    (\Im z(\z,u))\cdot  (\Im z(\z,u)) = \frac {\y^2_M}{u^2}>0; 
\label{inclu}
\end{eqnarray}
equation~(\ref{2.5.gg}) then implies that the Minkowskian tubes $T_+$ and $T_-$ of the leaves are contained within $\mathcal{Z}_+$ and, respectively, $\mathcal{Z}_-$ (see \cite{bem} for further details). This property also has important technical significance: in particular, it allows the computation of Fourier transforms of the waves with respect to the coordinates $\mathbf{x}^\mu$, which we will carry out in the following sections.

Finally, a crucial geometrical observation is that the above partial complexification already encompasses the entire Euclidean manifold $EAdS_d$.

In the following we will consider the  submanifolds of the null cone defined by the condition 
\begin{equation}
\gamma_P^{\pm}= \left\{ \xi\in {\cal C}_d   ; \ \ \xi^{d-1}+\xi^{d}   = \pm 1 \right\}
\end{equation}
that we parametrize also by Lorentzian coordinates:
\begin{equation}
 \xi(\eta)=\left\{\begin{array}{lcc}
\xi^{\mu} &=&  {\eta^\mu}\\
\xi^{d-1} &=&  \pm \frac12 (1+ {\eta}^2)    \\
\xi^{d} &=& \pm \frac12 (1- {\eta}^2)
\end{array} \right.
\label{conecoor0}.
\end{equation}
The invariant measure on $\gamma_P^{(\pm)}$ is 
$ d\mu(\xi(\eta)) = d\eta $
and the scalar product ${x \cdot \xi}$ 
takes the forms
\begin{eqnarray}
&&x(t,u) \cdot \xi(\eta)=
 \frac{u}{2}-\frac{\left({\x}-{\eta}\right)^2}{2u}, \ \  \ \ \ \  \xi(\eta)\in \gamma_P^{(+)}, \label{explicit}
\\
&& x(t,u) \cdot \xi(\chi)=
- \frac{u}{2}+\frac{\left({\x}+{\chi}\right)^2}{2u}, \ \  \ \  \xi(\chi)\in \gamma_P^{(-)}.\label{explicit2}
 \end{eqnarray}

\subsection{Integration cycles} \label{poincare}
Let us  specialize the first point in the two-point function (\ref{pwdu}) at the basis point $z_1= e_d= (0,\ldots,0,1)$. In spacetime dimension $d>2$ the  cycle $\gamma_S(e_d)$ (see Eq. (\ref{cycle00d}))
is  homeomorphic to a truncated cylinder with spherical section 
\begin{equation}
    \gamma_S(e_d)\cong [-\pi/2,\pi/2]\times S_{d-2}
\end{equation} 
and its  boundary is the disjoint union of two spheres
\begin{eqnarray}
&& \partial \gamma_S(e_d) =  \left\{(\xi^{0})  = \pm 1 , \ \  (\xi^{1})^2+\ldots +({\xi^{d-1}})^2 =1\right\}.
\end{eqnarray}
In Poincar\'e coordinates we would rather consider the following relative cycle contained in $\gamma_P^{+}$:
\begin{eqnarray}
&&  \gamma_P^{+}(e_d) =  \left\{ \xi\in \gamma^{+}_P :\ \  e_d\cdot \xi(\eta)= \frac12 (1- {\eta}^2) \geq 0\right\}.
\end{eqnarray}
The boundary of $ \gamma_P^{+}(e_d)$, namely  its intersection with the plane $\xi^d=0$, is the disjoint union of two Lobatchevski spaces:
\begin{equation}
\partial \gamma_P^{+}(e_d) \rightarrow \xi(\eta)=\left\{\begin{array}{lcc}
\xi^{0} &=&  \pm \sqrt{1+\vec \eta^2}\\
\xi^{i} &=&  {\eta^i}\\
\xi^{d-1} &=&  1   \\
\xi^{d} &=& 0
\end{array} \right. .
\label{conecoor2}
\end{equation}
 $\gamma_P^{+}$ is indeed mapped into $\gamma_S$  by the local rescaling
\begin{equation}
\Theta ( \xi(\eta)= \left\{\begin{array}{lcc}
\xi^{\mu} &=&  \frac{{2\eta^\mu}}{\sqrt{1+2{\eta^0}^2+2\vec\eta^2+(\eta^2)^2 }}\\ 
\xi^{d-1} &=&  \frac{ (1+ {\eta}^2)}{{\sqrt{1+2{\eta^0}^2+2\vec\eta^2+(\eta^2)^2 }}}    \\
\xi^{d} &=& \frac{ (1- {\eta}^2)}{{\sqrt{1+2{\eta^0}^2+2\vec\eta^2+(\eta^2)^2 }}}
\end{array} \right. .
\label{conecoor21}
\end{equation}
The  image  of $\partial \gamma_P^{(+)}(e_d)$ under the above map is the disjoint union of two open half-spheres:  
\begin{eqnarray}
    \Theta( \partial\gamma_P^{(+)}(e_d))= \left\{\pm 1,\frac{\vec\eta}{\sqrt{\vec{\eta}^2+1}},\frac{1}{\sqrt{\vec\eta^2+1}},0\right\}; 
\end{eqnarray} 
points of the equator are attained in the  limit  $\vec\eta \to \infty$ and correspond to points at infinity of the Lobachevsky space. 

Points of $\Theta(\gamma_P^{+}(e_d))$ have positive $(d-1)$-coordinate;  
the image of the cycle $\gamma_P^{+}(e_d)$ under the mapping $\Theta$ is therefore homeomorphic to a cylinder whose basis is a half-sphere; to build a cycle equivalent to $\gamma_S(e_d)$ we must add to  $\gamma_P^{+}(e_d)$ a second cycle contained in $\gamma_P^{-}$, namely 
\begin{eqnarray}
&&  \gamma^{-}_P(e_d) =  \left\{ \xi\in \gamma_P^- :\ \  e_d\cdot \xi(\chi)= -\frac12 (1- {\chi}^2) \geq 0\right\}.
\end{eqnarray}
It is immediately seen that
\begin{eqnarray}
    \Theta[ \partial \gamma^{-}_P(e_d)]= \left\{\pm 1,\frac{\vec\chi}{\sqrt{\vec{\chi}^2+1}},-\frac{1}{\sqrt{\vec\chi_2^2+1}},0\right\}
\end{eqnarray} 
is the disjoint union of the two lacking  halves of the spheres of the boundary of $\gamma_S(e_d)$.
\section{Waves and two-point functions in Poincar\'e  coordinates: study of the two dimensional case}\label{sec13}
\begin{figure}
            \centering
        \includegraphics[width=0.7\linewidth]{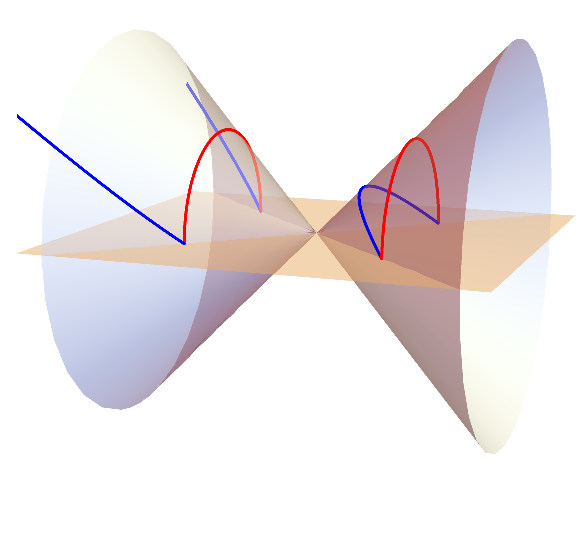}
                                \caption{For a given real event $x$ all the spherical and parabolical  integration cycles are shown; they have their endpoints on the plane $\xi\cdot x=0$ and belong to the same relative homology class. In the special case $d=2$ it is enough to integrate either one of the red or the blue  t cycles which are disconnected.}
        \label{fig:1}
\end{figure}

The two-dimensional case is simpler but still provides a valuable training ground for tackling the more general and challenging higher-dimensional case.

In spacetime dimension $d=2$, the leaves $\Pi_u$ are time-like parabolae, each parametrized by a single time coordinate $\x^0=t$, i.e.,
\begin{equation}
x^0 = \frac{t}{u} \,,
\end{equation}
(see Eq.~\ref{coordinates}). Similarly, $\gamma^+_P$ is parametrized by a single dual variable $\eta^0 = \eta$, which is conjugate to the time $t$.

 Let us chose the first point $z_1=x_1$ in Eq. (\ref{pw}) on the real AdS manifold (on the boundary of the backward tube);  $x_1$ has real coordinates $(t,u)$; the second point $z_2$ is taken in the interior of  the forward tube ${\cal Z}_+$ and is parametrized by the coordinates $(t'+i s' , u')$ with $s'$ positive. 
Our general formula requires that we integrate on a path homologous to the cycle 
\begin{equation}
   \gamma_P^{+}(x_1) =  \left\{ \xi\in \gamma^{+}_P :\ \  x_1(t,u)\cdot \xi(\eta)= \frac{u}{2}-\frac{\left({t}-{\eta}\right)^2}{2u} \geq 0\right\}.
\end{equation}
The result is
\begin{eqnarray}
&& W^{(2)}_\lambda(x_1(t,u),z_2(t'+is',u') )=
\frac{1}{2\pi }Q_\lambda\left(\frac{u^2+{u'}^2-\left(t-t'-i s'\right)^2}{2 u u'}\right) =\cr&& =
\frac{1}{4\pi}\int_{t-u}^{t+u}\left(\frac{u^2-(t-\eta )^2}{2 u}\right)^{\lambda} \left(\frac{{u'}^2-(t'+i s'-\eta )^2}{2 u'}\right)^{-\lambda-1}
d\eta. \label{1010a}
\end{eqnarray}
This provides yet another integral representation of the Legendre function of the second kind, $Q_\lambda$, which is naturally associated with the Poincar\'e coordinate system.

We now proceed to separate variables in the above expression by taking the Fourier transform of the holomorphic plane waves \eqref{waves} with respect to the time coordinate $t$. The result is
\begin{eqnarray}
&& \int (x(t - i\epsilon, u) \cdot \xi(\eta) )^\lambda e^{i \omega t } \, dt 
\,=\, \theta(\omega) \, \frac{(2 \pi)^{\frac{3}{2}} u^{\frac{1}{2}}}{\Gamma(-\lambda)} 
\, e^{i \omega \eta} \, \omega^{-\frac{1}{2} - \lambda} \, 
J_{-\frac{1}{2} - \lambda}(\omega u), \\
&& \int (x(t + i\epsilon, u) \cdot \xi(\eta) )^\lambda e^{i \omega t } \, dt 
\,=\, \theta(-\omega) \, \frac{(2 \pi)^{\frac{3}{2}} u^{\frac{1}{2}}}{\Gamma(-\lambda)} 
\, e^{i \omega \eta} \, |\omega|^{-\frac{1}{2} - \lambda} \, 
J_{-\frac{1}{2} - \lambda}(|\omega| u), \label{po}
\end{eqnarray}
valid for $-1 < \Re \lambda < 0$, where $\theta(\omega)$ is Heaviside's step function.  

A physically relevant feature of the above Fourier transforms is that plane waves holomorphic in the backward tube of the complex AdS manifold contain only positive frequencies with respect to the time coordinate of the Poincar\'e branes, whereas plane waves holomorphic in the forward tube contain only negative frequencies. This support property follows from the geometric inclusion of the tubes described in Eq.~\eqref{inclu}.  

Inversion then allows us to rewrite the integral \eqref{1010a} as follows:
\begin{eqnarray}
&& \int\left( \frac{u}{2}-\frac{\left(t-{\eta}\right)^2}{2u
}\right)^{\lambda} \left( \frac{u'}{2}-\frac{\left({t'+is'}-{\eta}\right)^2}{2u}\right)^{-\lambda-1} d\eta=
  \cr &&
=\frac{(2 \pi u' )^{1/2}}{\Gamma (\lambda+1 )}\int \int_{-t-u}^{{t+u}}
\left( \frac{u}{2}-\frac{\left(t-{\eta}\right)^2}{2u
}\right)^{\lambda} e^{-i \omega(t'+i s'-\eta)} { |\omega| ^{\frac{1}{2}+\lambda }J_{\frac{1}{2} + \lambda }(| \omega| 
   u') \theta (-\omega)} d\omega d\eta.  \cr &&
\end{eqnarray}
The integral over $\eta$ contains both positive and negative frequencies (no step function is present here): 
\begin{eqnarray}
 \int_{t-u}^{{t+u}}
\left( \frac{u}{2}-\frac{\left(t-{\eta}\right)^2}{2u
}\right)^{\lambda} e^{i \omega\eta} 
d\eta  = \sqrt{2 \pi u} |\omega|^{-\lambda -\frac{1}{2}} \Gamma (\lambda +1) J_{\frac{1}{2} + \lambda }(|\omega|u), \ \ \ \Re \lambda>-1. \label{16}
\end{eqnarray}
Putting everything together, we arrive at the final result in the form of a Hankel representation: 
\begin{eqnarray}
W^{(2)}_\lambda\big(x_1(t,u), z_2(t'+i s', u')\big) 
&=& \frac{1}{2} \, (u u')^{1/2} \int_0^\infty 
e^{-i \omega (t - t' - i s')} \, 
J_{\frac{1}{2} + \lambda}(\omega u) \, 
J_{\frac{1}{2} + \lambda}(\omega u') \, d\omega, \cr &&
\label{24} 
\end{eqnarray}
which has been proven under the condition $-1 < \Re \lambda < 0$, but is valid more generally in the sense of distributions. It provides a diagonalization of the two-point function which, for real $\lambda$, manifestly exhibits its positive-definiteness in the region where it converges.  

By comparing this expression with the one obtained by inserting the Fourier representation of both the first and second factors in Eq.~\eqref{1010a}, we also recover a classical formula of Sonine, which we do not reproduce here \cite{watson}.

\subsection{The other cycle}

In preparation for the general case, it is instructive to integrate also over the other non-compact parabolic cycle (see Figure~\ref{fig:1}). Even though we expect a result identical to \eqref{24}, the method to reach it is rather different, as the integration cycle is now unbounded and extends to infinity. The delicate point lies in the behavior of the waves at infinity, and the strategy to circumvent this difficulty will provide crucial guidance for handling the problem in general spacetime dimension $d$.

The relevant relative cycle is 
\begin{eqnarray}
    \gamma_P^{-}(x_1)= \left\{ \xi\in \gamma_P^- :\ \  x_1\cdot \xi(\chi)=- \frac{u}{2}+\frac{\left({\x}+{\chi}\right)^2}{2u} \geq 0\right\}.
\end{eqnarray}
This choice leads to an alternative representation of the two-point function:
\begin{eqnarray}
   W^{(2)}_\lambda(x_1,z_2)      =\frac 1{4\pi}\int_{-\infty} ^{-t-u}+\int_{-t+u} ^{\infty}\left(\frac{(t+\eeta )^2-u^2}{2u }\right)^\lambda   \left(\frac{(t'+\eeta )^2-{u'}^2}{2u' }\right)^{-\lambda -1}  d\eeta
          \label{28}
\end{eqnarray}
where it is understood that
\begin{eqnarray}
    \lim_{\eeta\to\pm \infty}\text{Arg}\left(\frac{(t'+\eeta )^2-{u'}^2}{2u' }\right)  = 0. \label{phase}
\end{eqnarray}
This condition is to be interpreted as follows.  
In formula~\eqref{28}, there are two disconnected semi-infinite intervals of integration; the phase of the second factor of the integrand must be chosen with two different determinations, one for each connected part of the integration cycle.
Equation~\eqref{28} thus actually means
\begin{eqnarray}
    && W^{(2)}_\lambda(z_1,z_2) 
     =\frac 1{4\pi}\int_{-\infty} ^{-t-u}\left(\frac{(t+\eeta )^2-u^2}{2u }\right)^\lambda   \left(\frac{(t'+\eeta )^2-{u'}^2}{2u' }\right)_-^{-\lambda -1}  d\eeta
     \cr && + \frac 1{4\pi}\int_{-t+u} ^{\infty}\left(\frac{(t+\eeta )^2-u^2}{2u }\right)^\lambda   \left(\frac{(t'+\eeta )^2-{u'}^2}{2u' }\right)_+^{-\lambda -1}  d\eeta
      \label{29}
\end{eqnarray}
with the two distinct determinations ($\pm$) are specified at plus and minus infinity by  Eq. (\ref{phase}). The Fourier transforms of the waves according to the above determinations are given by  
\begin{eqnarray}&& \int \left(\frac{( t' +i s' +\eeta)^2-{u'}^2}{2u' }\right)_\pm^{-\lambda -1} e^{i\omega t'}  dt'  = \cr && =  \pm i \theta(-\omega)  e^{-i \omega (\chi+i s')} | \omega| ^{\frac{1}{2}+\lambda } \left(e^{\mp 2 i \pi  \lambda }-1\right) \sqrt{2 \pi u'} J_{\frac{1}{2}+\lambda }\left(|  \omega |  u'\right) \Gamma (-\lambda ).
\end{eqnarray}
Inversion gives the Fourier representations of the plane waves:
\begin{eqnarray}&& \left(\frac{(t' +i s' +\eeta)^2-{u'}^2}{2u' }\right)_\pm ^{-\lambda -1} = \cr&& = \pm \frac{i  \Gamma (-\lambda ) \sqrt{u'} \left(e^{\mp 2 i \pi  \lambda }-1\right) }{\sqrt{2 \pi }} \int_{-\infty}^0 e^{-i\omega (t'+is'+\eeta)} { | \omega | ^{\frac{1}{2}+\lambda }  J_{\frac{1}{2}+\lambda }\left(| \omega |  u'\right)}d\omega.
\end{eqnarray}
Inserting these formulae into (\ref{29}) we get
\begin{eqnarray}
    && W^{(2)}_\lambda(z_1,z_2) = \cr &&
     \cr &=&     \frac { \Gamma (-\lambda ) \sin(\pi \lambda)\sqrt{u'}}{4\pi \sqrt{2 \pi }}\int_{-\infty}^0 e^{i\omega (t-t' -i s')}   { | \omega | ^{\frac{1}{2}+\lambda }   J_{\frac{1}{2}+\lambda }\left(| \omega |  u'\right)}{}d\omega \times \cr &&{}\  \  \  \ \ \ \ \ \ \ \ \ \ \  \times \int_{u} ^{\infty}  \cos(\omega \eeta-\pi\lambda)\left(\frac{\eta ^2-u^2}{2u }\right)^\lambda    d \eeta =
\cr &=&   \frac 12 {( u u' )^{1/2}}\int_0^\infty e^{-i \omega(t-t' -i s')} J_{\frac{1}{2}+\lambda}(\omega u)J_{\frac{1}{2} + \lambda }(\omega 
   u') d\omega .
\end{eqnarray}
As anticipated, we recover the same Hankel representation \eqref{24}, although by a significantly more involved procedure. In general spacetime dimension $d>2$, the two separate integration cycles contribute differently. Therefore, it is essential to account for both contributions in order to obtain the complete result.

\section{Waves and two-point functions in Poincar\'e  coordinates: general case}
\label{sec14}

Let us now face the problem of diagonalizing the two-point function \eqref{pwduint} in Poincar\'e coordinates in arbitrary spacetime dimension $d$. To this end, we choose 
\[
x_1 = x_1(\x, u)\in  AdS_M \quad \text{and} \quad z_2 = z_2(\z', u') \in \ZZ_+, \quad \z' \in T_+,
\] 
in the relevant complex Minkowskian brane.  
The integration cycle consists of two disconnected parts, denoted as
\[
\gamma_P(x_1) = \gamma_P^+(x_1) \, \cup \, \gamma_P^-(x_1),
\]
where
\begin{eqnarray}
\gamma_P^+(x_1) &=& \left\{ \xi(\eta) \in \gamma_P^+ : (\eta^0 - \x^0)^2 \leq (\vec{\eta} - \vec{\x})^2 + u^2 \right\}, \\
\gamma_P^-(x_1) &=& \left\{ \xi(\chi) \in \gamma_P^- : (\chi^0 + \x^0)^2 \geq (\vec{\chi} + \vec{\x})^2 + u^2 \right\}.
\end{eqnarray}

As emphasized previously, unlike the two-dimensional case, it is essential to integrate over both disconnected cycles. The following expression is the one that requires evaluation:
\begin{eqnarray}
&& W^{(d)}_\lambda(x_1(\x,u),z_2(\z',u') )
=\cr&& \cr&& =
c_d(\lambda)\int d\vec \eta \int_{x^0-\sqrt{\left(\vec \eta-\vec\x\right)^2 + u^2}}^{x^0+\sqrt{\left(\vec \eta-\vec\x\right)^2 + u^2}}\left(\frac{u^2-(\x-\eta )^2}{2 u}\right)^{\lambda} \left(\frac{{u'}^2-(\z'-\eta )^2}{2 u'}\right)^{-\lambda-d+1}
d\eta^0 \cr && +
c_d(\lambda)\int d\vec\eeta\int_{-\infty}^{-\x^0-\sqrt{\left(\vec \chi+\vec\x\right)^2+u^2}} \left(\frac{(\x+\eeta )^2-u^2}{2u }\right)^\lambda   \left(\frac{(\z'+\eeta )^2-{u'}^2}{2u' }\right)^{-\lambda-d +1}  d\eeta^0 +
\cr && +
c_d(\lambda)\int d\vec\eeta\int^{\infty}_{-\x^0+\sqrt{\left(\vec \chi+\vec\x\right)^2+u^2}} \left(\frac{(\x+\eeta )^2-u^2}{2u }\right)^\lambda   \left(\frac{(\z'+\eeta )^2-{u'}^2}{2u' }\right)^{-\lambda-d +1}  d\eeta^0
\label{long}
\end{eqnarray}
Here, the first line corresponds to the integral over $\gamma_P^+(x_1)$, while the second and third lines correspond to the integral over $\gamma_P^-(x_1)$. The challenging task of evaluating these three integrals will be rewarded by an exceptionally simple final result. Moreover, the calculation will yield new integral formulae involving Bessel functions.

\vskip 5 pt
Let us start with the integral over $\gamma_P^+(x_1)$. In this case, the relevant Fourier transform of the waves is given by
\begin{eqnarray}
\widetilde{\phi^\pm_{\lambda}}(\k) &=& \int \big(x(\x^0 \pm i \epsilon, \vec{\x}, u) \cdot \xi(\eta) \big)^\lambda \, e^{i \k \cdot \x} \, d\x \\
&=& e^{i \k \cdot \eta} \int \left[ \frac{u}{2} - \frac{(\y^0 \pm i \epsilon)^2 - \vec{\y}^2}{2 u} \right]^\lambda e^{i \k \cdot \y} \, d\y,
\end{eqnarray}
where $\y = \x - \eta$, and we leave the dependence on $u$ of the left-hand side implicit.  

Let us choose $\k = (\k^0, {0})$ and perform the radial integration first:
\begin{eqnarray}
\widetilde{\phi^+_{\lambda}}(\k^0, 0) 
&=& e^{i \k^0 \eta^0} \frac{2 \pi^{\frac{d-2}{2}}}{\Gamma\left(\frac{d-2}{2}\right)} 
\int_{-\infty}^{\infty} d\y^0 \int_0^\infty \left[ \frac{u}{2} - \frac{(\y^0 + i \epsilon)^2 - r^2}{2 u} \right]^\lambda e^{i \k^0 \y^0} r^{d-3} \, dr \\
&=& \frac{(2u)^{-\lambda} \, e^{i \k^0 \eta^0} \, \pi^{\frac{d-2}{2}} \, \Gamma\left( \frac{2-d}{2} - \lambda \right)}{\Gamma(-\lambda)}
\int_{-\infty}^{\infty} \left( u^2 - (\t + i \epsilon)^2 \right)^{\frac{d-2}{2} + \lambda} e^{i \k^0 \t} \, d\t \label{ugo}
\end{eqnarray}
which is valid for $\Re(2\lambda + d) < 2$ and $\Re d > 2$. These restrictions will be removed at the end when the integral will be considered in the sense of distributions.

When $\k^0 > 0$, the contour of integration can be pushed to infinity, and the integral is then seen to vanish. The geometric and physical reason behind this possibility is, once again, the previously mentioned immersion of the Minkowskian tubes of the branes into the AdS chiral tuboids.

When 
$\k^0 <0 $ the above contour distortion is not possible and we get 
\begin{eqnarray}
 \widetilde{\phi^+_{\lambda}}(\k^0,0)   
  = \frac{(2\pi)^{\frac{d+1}{2} }}{\Gamma \left(-\lambda \right)} \theta(-\k^0)  {|\k^0|}^{-\frac{d-1}{2}-\lambda }
   e^{i\k^0 \eta^0}u^{\frac{d-1}{2}}J_{-\frac{d-1}{2}-\lambda }(| \k^0 |  u). \label{45}
\end{eqnarray}
  For what follows it is also useful to take in Eq. (\ref{ugo})  the $\t$ integration first: 
\begin{eqnarray}
\widetilde{\phi^+_{\lambda}}(\k^0,0) 
=\frac{4 \sqrt{2} | \k^0 | ^{-\frac{1}{2}-\lambda } e^{i \eta  \k^0 } \pi ^{\frac{d+1}{2}} 
   u^{-\lambda } }{\Gamma \left(\frac{d-2}{2}\right) \Gamma (-\lambda )} \int_0^\infty r^{d-3} \left(r^2+u^2\right)^{\frac{1}{4}+\frac{\lambda }{2}} J_{-\frac{1}{2} - \lambda }\left(| \k^0 | 
   \sqrt{r^2+u^2}\right). \label{49}
\end{eqnarray}
By comparing Eqs.~\eqref{45} and \eqref{49}, we deduce the identity
\begin{equation}
\int_0^\infty r^{d-3} \left(r^2 + u^2\right)^{\frac{1}{4} + \frac{\lambda}{2}} 
J_{-\frac{1}{2} - \lambda}\Big(|\k^0| \sqrt{r^2 + u^2}\Big) \, dr
= 2^{\frac{d-4}{2}} \Gamma\Big(\frac{d-2}{2}\Big) |\k^0|^{ \frac{2-d}{2}} u^{\frac{d-1}{2} + \lambda} 
J_{-\frac{d-1}{2} - \lambda}(|\k^0| u), 
\label{Wa}
\end{equation}
valid when the integral converges. This identity is a special case of a known integral formula by Watson \cite{watson} and will be used later.

Restoring Lorentz invariance, we obtain the Fourier integral representation of the wave (written in a form ready to be used in our calculation):
\begin{eqnarray}
&& (x(\z', u') \cdot \xi(\eta) )^{-\lambda-d+1}=  \left(\frac{{u'}^2-(\z' -\vec\eta)^2}{2 u'}\right)^{-\lambda-d+1} =  \cr && =\frac{(2 \pi )^{\frac{3-d}{2}} {u'}^{\frac{d-1}{2}}}{\Gamma (\lambda +d-1)}   \int 
 e^{-i \k (\z'-\eta)  } \theta(-\k^0) \theta(\k^2) \left( \sqrt{\k^2} \right) ^{\frac{d-1}{2} +\lambda }
   J_{\frac{d-1}{2} +\lambda }(\sqrt{\k^2}\,   u') d\k,  \cr && \label{88}
\end{eqnarray}
where the factor $\theta(-\k^0)$ guarantees convergence of the integral as long as $\z' \in T_+$.

We are now ready to evaluate the first line of Eq.~\eqref{long} (i.e., the integral over the cycle $\gamma_P^+(x_1)$). Without loss of generality, we may simplify the calculation by choosing $\x = 0$, i.e.,
\[
x_1 = x_1(0, u).
\]
Proceeding as in the two-dimensional case (see Eq.~\eqref{16}), we are led to examine the integral
\begin{eqnarray}
&& \frac{2 \pi^{\frac{d-2}{2}}}{\Gamma\left(\frac{d-2}{2}\right)} 
\int_0^\infty r^{d-3} \, dr 
\int_{-\sqrt{r^2 + u^2}}^{\sqrt{r^2 + u^2}} e^{i \k^0 \eta^0} 
\left(\frac{u^2 + r^2 - (\eta^0)^2}{2u}\right)^\lambda d\eta^0 =\nonumber\\
&&= \frac{2 \sqrt{2} \pi^{\frac{d-1}{2}} |\k^0|^{-\lambda - \frac{1}{2}} \Gamma(\lambda + 1) u^{-\lambda}}{\Gamma\left(\frac{d-2}{2}\right)} 
\int_0^\infty r^{d-3} \left(r^2 + u^2\right)^{\frac{\lambda}{2} + \frac{1}{4}} 
J_{\lambda + \frac{1}{2}}\left(|\k^0| \sqrt{r^2 + u^2}\right) dr. \cr&& 
\label{57a}
\end{eqnarray}
Here we set $|\vec{\eta}| = r$ and integrate over the angular variables, assuming that $\Re \lambda > -1$. Although the integral on the right-hand side of (\ref{57a}) is similar to that in (\ref{49}), it does not appear to be directly known. Following the procedure used in Eq.~(\ref{49}), we evaluate it by first integrating over the variable $r$ on the left-hand side of (\ref{57a}):
\begin{eqnarray}
&&  \frac{2 \pi ^{\frac{d-2}{2}}}{\Gamma \left(\frac{d-2}{2}\right)} \int_0^\infty r^{d-3}  dr  \int_{-\sqrt{r^2 + u^2}}^{\sqrt{r^2 + u^2}} e^{i\k^0 \eta^0}   \left(\frac {u^2+r^2-{\eta^0}^2}{2u} \right)^\lambda  d\eta^0 =
\cr&&=  -\frac{(2\pi u) ^{\frac{d-1}{2}}  |\k^0|^{-\frac{d-1}{2}-\lambda}
 \Gamma (\lambda +1)}{  \cos \left(\frac{1}{2} \pi  (d+2 \lambda )\right)} \left(\cos (\pi  \lambda ) J_{\frac{d-1}{2}+\lambda
   }(u |\k^0| )-\sin \left(\frac{\pi  d}{2}\right) J_{-\frac{d-1}{2} - \lambda }(u
   |\k^0 |)\right). \cr && \label{op99}
\end{eqnarray}
{Comparison  gives the following result (valid where the integral converges):
\begin{eqnarray}
 && 
\int_0^\infty r^{d-3} \left(r^2+u^2\right)^{\frac{\lambda }{2}+\frac{1}{4}} J_{\lambda
   +\frac{1}{2}}\left(|\k^0 | \sqrt{r^2+u^2} \right) dr = \cr &&  \cr&&=  \frac{2^{\frac{d-4}{2}} \pi \,  | \k^0 | ^{-\frac{d-2}{2}} u^{\frac{d-1}{2} + \lambda }}{\cos \left(\frac{1}{2} \pi  (d+2
   \lambda )\right) \Gamma \left(2-\frac{d}{2}\right)}\left(\frac{\cos (\pi  \lambda )}{ \sin
   \left(\frac{d \pi }{2}\right) }J_{\frac{d-1}{2} +\lambda }(| \k^0 |  u)- J_{-\frac{d-1}{2}- \lambda }(| \k^0 |  u)\right).\label{55}
\end{eqnarray}}
 Restoring Lorentz invariance in Eq. (\ref{op99})  gives 
 \begin{eqnarray}
  \int   \int_{-\sqrt{u^2+\vec \eta^2}}^{\sqrt{u^2+\vec \eta^2}}\left(\frac{u^2+\vec\eta^2-{\eta^0} ^2}{2 u}\right)^{\lambda} e^{i \k \eta}
d\eta^0 d\eta 
=  -\frac{(2\pi u) ^{\frac{d-1}{2}}   \left(\sqrt{\k^2} \right)^{-\frac{d-1}{2}-\lambda}
 \Gamma (\lambda +1)}{  \cos \left(\frac{1}{2} \pi  (d+2 \lambda )\right)} \times\cr \times \theta(\k^2) \left(\cos (\pi  \lambda ) J_{\frac{d-1}{2}+\lambda
   }\left(u \sqrt{\k^2} \right)-\sin \left(\frac{\pi  d}{2}\right) J_{-\frac{d-1}{2} - \lambda }\left(u
    \sqrt{\k^2} \right)\right). \label{87}
   \end{eqnarray}
By substituting Eqs.~(\ref{88}) and (\ref{87}) into Eq.~(\ref{long}), we obtain the first part of the result, computed for the special case in which $x_1 = x_1(0,u)$:
\begin{eqnarray}
 \int   \int_{-\sqrt{u^2+\vec \eta^2}}^{\sqrt{u^2+\vec \eta^2}}\left(\frac{u^2+\vec\eta^2-{\eta^0} ^2}{2 u}\right)^{\lambda} \left(\frac{{u'}^2(\z' -\eta)^2}{2 u'}\right)^{-\lambda-d+1}
d\eta^0 d\eta =  \cr =  -    \frac{ 2 \pi ({u u'})^{\frac{d-1}{2}}\Gamma (\lambda +1) }{\Gamma (\lambda +d-1) \cos \left(\frac{1}{2} \pi  (d+2 \lambda )\right)} \int e^{-i \k\z'  } \theta(-\k^0) \theta(\k^2) \frac{  
}{  } \times\cr \times \left(\cos (\pi  \lambda ) J_{\frac{d-1}{2}+\lambda
   }(u \sqrt{\k^2} )-\sin \left(\frac{\pi  d}{2}\right) J_{-\frac{d-1}{2} - \lambda }(u
    \sqrt{\k^2} )\right)   
   J_{\frac{d-1}{2} +\lambda }(u \sqrt{\k^2}) d\k.\label{57}
   \end{eqnarray}

Let us now focus on the other cycle $\gamma^{(-)}_P(x_1)$. In this case the situation is more complicated because the cycle passes through infinity and, as in Eq.~(\ref{29}), we need the Fourier transforms of the waves whose phase vanishes at infinity. The analogue of Eq.~(\ref{phase}) is here
\begin{eqnarray}
    \lim_{\chi^0\to\pm \infty}\text{Arg}\left(\frac{(\x'+\chi )^2-{u'}^2}{2u' }\right)  = 0.
\end{eqnarray}
 Proceeding as before we compute the relevant Fourier transforms of the waves with the correct phases at infinity: 
 \begin{eqnarray}
  \left(\frac{(\x'+\chi )^2+{u'}^2}{2 u'}\right)_\pm^{-\lambda-d+1}  =\pm {i (2 \pi )^{\frac{1-d}{2}}\left(e^{\mp 2 i \pi  (d+\lambda )}-1\right) }{\Gamma (2-\lambda -d)}  \times \cr  \times {u'}^{\frac{d-1}{2}}  \int 
 e^{-i \k (\x'+\chi)  }\theta(-\k^0) \theta(\k^2) \left( \sqrt{\k^2} \right) ^{\frac{d-1}{2} +\lambda }
   J_{\frac{d-1}{2} +\lambda }(\sqrt{\k^2}\,   u') d\k . \label{88o}
\end{eqnarray}
In the last step  we evaluate the integral 
\begin{eqnarray}
 && 
 \frac{2 \pi ^{\frac{d-2}{2}}}{\Gamma \left(\frac{d-2}{2}\right)} \int_0^\infty r^{d-3}  dr  \int_{\sqrt{r^2 + u^2}}^{\infty} e^{-i \k^0 \chi^0}   \left(\frac {{\chi^0}^2 -u^2-r^2}{2u} \right)^\lambda  d\chi^0 =\cr&&
 = -\frac{\sqrt{2} \, \pi ^{\frac{d-1}{2}} \Gamma (\lambda +1) u^{-\lambda } | \k^0| ^{-\lambda -\frac{1}{2}}
    }{\Gamma \left(\frac{d}{2}-1\right)} \times 
    \cr&& \int_0^\infty r^{d-3}  dr\left(r^2+u^2\right)^{\frac{\lambda }{2}+\frac{1}{4}}\left( \frac{J_{\lambda
   +\frac{1}{2}}\left(| \k^0|\sqrt{r^2+u^2}  \right)}{\cos (\pi  \lambda )}+(\tan (\pi  \lambda )-i) J_{-\lambda -\frac{1}{2}}\left( |
   \k^0|\sqrt{r^2+u^2} \right)\right) = 
   \cr&& 
   = 
\frac { (2\pi u)^{\frac{d-1}{2}} \Gamma (\lambda +1) | \k^0 | ^{-\frac{d-1}{2} -\lambda }}{2}
   \left(\frac{ J_{\frac{d-1}{2}+\lambda }(u | \k^0 | )}{\cos \left(\frac{\pi  d}{2}+\pi  \lambda \right)}-\left(\tan
   \left(\frac{\pi  d}{2}+\pi  \lambda \right)-i\right) J_{-\frac{d-1}{2} - \lambda }(u | \k^0 | )\right) \cr && 
\end{eqnarray}
where we used both Eqs. (\ref{49}) and (\ref{55}); as before  we  have assumed $\Re\lambda>-1$ and  set $\k=(\k^0,0)$ (with $\k^0<0$).
Similarly 
\begin{eqnarray}
 && 
 \frac{2 \pi ^{\frac{d-2}{2}}}{\Gamma \left(\frac{d-2}{2}\right)} \int_0^\infty r^{d-3}  dr  \int^{-\sqrt{r^2 + u^2}}_{-\infty} e^{-i \k^0 \chi^0}   \left(\frac {{\chi^0}^2 -u^2-r^2}{2u} \right)_-^\lambda  d\chi^0 =\cr&& =
\frac {  (2\pi u)^{\frac{d-1}{2}} \Gamma (\lambda +1) | \k^0 | ^{-\frac{d-1}{2} -\lambda }}{2}
   \left(\frac{ J_{\frac{d-1}{2}+\lambda }(u | \k^0 | )}{\cos \left(\frac{\pi  d}{2}+\pi  \lambda \right)}-\left(\tan
   \left(\frac{\pi  d}{2}+\pi  \lambda \right)+i\right) J_{-\frac{d-1}{2} + \lambda }(u | \k^0 | )\right). \cr && 
\end{eqnarray}
Taking into account Eq.(\ref{88o}) we get
\begin{eqnarray}
 && 
\int_{\gamma^-_P(x_1) }
\left(\frac {{\chi^0}^2 -u^2-\vec\chi^2}{2u} \right)
\left(\frac{(\z'+\chi )^2-{u'}^2}{2 u'}\right)_-^{-\lambda-d+1}  d\chi \, +  \cr &+& \int_{\gamma^+_P(x_1) }
\left(\frac {{\chi^0}^2 -u^2-\vec\chi^2}{2u} \right)
\left(\frac{(\z'+\chi )^2-{u'}^2}{2 u'}\right)_+^{-\lambda-d+1} d\chi = \cr &=&
\frac{2\, \Gamma (\lambda +1) \sin (\pi  (d+\lambda ))  \Gamma (-d-\lambda +2)
   \left(u u'\right)^{\frac{d-1}{2}} }{\cos \left(\frac{\pi  d}{2}+\pi  \lambda \right)}  \int e^{-i \k\z  } \theta(-\k^0) \theta(\k^2) \times \cr && 
\times  J_{\frac{d-1}{2}+\lambda }\left(  u'\sqrt{\k^2}\right) \left(\cos (\pi  (d+\lambda ))
   J_{\frac{d-1}{2}+\lambda }\left(  u\sqrt{\k^2}\right)+\sin
   \left(\frac{\pi  d}{2}\right) J_{-\frac{d-1}{2} - \lambda }\left(  u\sqrt{\k^2}\right)\right). \cr&&  \label{yuppy}
\end{eqnarray}
By summing Eqs. (\ref{57}) and (\ref{yuppy}) and changing $\k$ to $-\k$ we obtain the diagonalization we were looking for:
\begin{eqnarray}
   &&  \int_{\gamma_P(x_1)} (x_1\cdot \xi)^\lambda (\xi\cdot z_2)^{-\lambda-d+1} d\mu_\gamma(\xi)=  \cr&=& 
    -\frac{4 \pi  \cos \left(\frac{\pi  d}{2}\right) \Gamma (\lambda +1) \left(u u'\right)^{\frac{d-1}{2}}}{\Gamma (d+\lambda
   -1)}  \int e^{-i \k(\x-\x')} \theta(\k^0) \theta(\k^2) J_{\frac{d-1}{2}+\lambda }(u \sqrt{k^2} ) J_{\frac{d-1}{2}+\lambda }\left(  u'\sqrt{k^2}\right)\cr && 
\end{eqnarray}
Finally we restore the correct normalization of the two-point function by multiplying both sides of the above formula by $c_d(\lambda)$ and get
\begin{eqnarray}
 && W^{(d)}_\lambda(z_1(\z_1,u),z_2(\z_2,u'))= \cr&&=
\frac{\left(u u'\right)^{\frac{d-1}{2}}}{2 (2\pi) ^{d-2} }
   \int e^{-i \k(\z_1-\z_2)} \theta(\k^0) \theta(\k^2) J_{\frac{d-1}{2}+\lambda }\left(u \sqrt{\k^2} \right) J_{\frac{d-1}{2}+\lambda }\left(  u'\sqrt{\k^2}\right)d\k. \label{JJ}
\end{eqnarray}
In the final step we write
\begin{eqnarray}
 \theta(\k^2) = \int_0^\infty \delta(\k^2-m^2) dm^2 \label{rell}
\end{eqnarray}
and insert Eqs. (\ref{tpm}) and (\ref{rell}) into Eq. (\ref{JJ}) to get 
\begin{equation}
 W^{(d)}_\lambda(z_1(\z_1,u),z_2(\z_2,u'))= 
\frac{1}{2  } \left(u u'\right)^{\frac{d-1}{2}}
   \int W^{M_{d-1}}_m(\z_1,\z_2) J_{\frac{d-1}{2}+\lambda }\left(m u  \right) J_{\frac{d-1}{2}+\lambda }\left(m  u'\right)dm^2. \label{JJ2}
\end{equation}
That  this expression  solves the Klein-Gordon equation
\begin{eqnarray}
     u^{2} \Box_M \phi- u^d{\partial_u}(   u^{2-d} \partial_u \phi ) +m^2_\lambda\phi = 0 
\end{eqnarray}
may be verified by using the  relation 
\begin{eqnarray}
   u^d{\partial_u}(   u^{2-d} \partial_u   \left(u^{\frac{d-1}{2}}J_{\frac{d-1}{2}+\lambda }\left(m u  \right) \right) = u^{\frac{d-1}{2}} \left(\lambda  (d+\lambda -1)-m^2 u^2\right) J_{\frac{d-1}{2}+\lambda
   }(m u). \label{jlk}
\end{eqnarray}

The remarkable Hankel formula (\ref{JJ2}) shows that the maximally analytic AdS two-point function can be represented as a K\"all\'en-Lehmann superposition of maximally analytic two-point functions defined in a  Minkowski space with one dimension less, explicitly linking the analytic and geometric structures of AdS and Minkowski spacetimes.

Eq.  (\ref{JJ2}) was anticipated  in \cite{tsi} and rediscovered by \cite{raju,raju2}. In  \cite{bertola1,bertola2} it was derived by employing Hilbert space methods suitable for generic warped manifolds; the value of the derivation in \cite{bertola1,bertola2} was also to clarify the Breitenlohner and Friedmann phenomenon \cite{breit} in terms of the self-adjointness of a one-dimensional transverse Schrodinger operator; however, a crucial part of the proof  was based on a known Hankel transform of the Legendre function $Q$ \cite{batemantransform2}. The present derivation is fully self-contained and specifically tailored to the geometry of the complex AdS manifold and the AdS spectral condition. 

Eq. (\ref{JJ2}) may be used to show the positive-definiteness of the two-point function. 
Let us briefly sketch how by introducing a sesquilinear form in the space ${\cal T}(AdS_M)$ of test function which are smooth and vanish outside $AdS_M$   as indicated in in Eq.  (\ref{4.5}):
\begin{eqnarray}
  &&  \langle f,f\rangle = \int W_\lambda^{(d)}(x_1,x_2)\overline {f(x_1)}f(x_2) \sqrt{g(x_1)} \sqrt{g(x_2)}dx_1 dx_2= \cr &&
    = \int d\k \frac{  \theta(\k^0) \theta(\k^2)}{2 (2\pi) ^{d-2} }
   \int \overline{\hat f(k,u)} \hat f(k,u')  \left(u u'\right)^{-\frac{d+1}{2}} J_{\frac{d-1}{2}+\lambda }\left(u \sqrt{\k^2} \right)J_{\frac{d-1}{2}+\lambda }\left(  u'\sqrt{\k^2}\right) {du} \ {du'} \cr &&
\end{eqnarray}
where 
\begin{eqnarray}
\hat f(\k,u) = \int e^{i \k\x} {f(x(\x,u) )} d\x
\end{eqnarray}
is the  Fourier transform the test function w.r.t the Poincar\'e coordinate $\x$. When  $\lambda$ is real the above expression becomes formally positive
\begin{eqnarray}
    \langle f,f\rangle =
 \frac{1}{2 (2\pi) ^{d-2} } \int
\theta(\k^0) \theta(\k^2) \left|\hat F_{\frac{d-1}{2}+\lambda}(\k)\right|^2 d\k. \label{JJ34}
\end{eqnarray}
where
\begin{eqnarray}
    \widehat F_{\frac{d-1}{2}+\lambda}(\k) =
 \int_0^\infty { u^{-\frac{d+1}{2}} }{\hat f(\k,u)}J_{\frac{d-1}{2}+\lambda }\left(u \sqrt{\k^2} \right) { du}
\end{eqnarray}
A thorough study of the Hilbert space structure will presented elsewhere. 

\section{Green functions, Schwinger functions  and Wick rotations}

\label{sec15}
Eq.  (\ref{JJ2})  may be used to promptly build  a new explicit Hankel representation of the Schwinger function (\ref{kgtps}) extended to non coincident Euclidean points of $EAdS_d$: 
\begin{equation}
S^{(d)}_\lambda(z_{\scriptscriptstyle{E}},
z'_{\scriptscriptstyle{E}})= 
\frac{1}{2  } \left(u u'\right)^{\frac{d-1}{2}}
   \int S^{E_{d-1}}_m(\x_{\scriptscriptstyle{E}},
\x'_{\scriptscriptstyle{E}}) J_{\frac{d-1}{2}+\lambda }\left(m u  \right) J_{\frac{d-1}{2}+\lambda }\left( m  u'\right)dm^2,  \label{JJG}
\end{equation}
where 
\begin{eqnarray}
 && S^{E_{d-1}}_m(\x_{\scriptscriptstyle{E}}) =  \frac{1}{(2\pi)^{d-1 }} \int   \frac{e^{-i\k_{\scriptscriptstyle{E}}\x_{\scriptscriptstyle{E}}}}{{\k_{\scriptscriptstyle{E}}^2+m^2}}  d\k_{\scriptscriptstyle{E}}
\end{eqnarray}
is the Schwinger function of a $(d-1)$-dimensional massive Euclidean scalar field and $\k_{\scriptscriptstyle{E}}\x_{\scriptscriptstyle{E}}= \k^0_{\scriptscriptstyle{E}}\x^0_{\scriptscriptstyle{E}}+\ldots +\k^{d-1}_{\scriptscriptstyle{E}}\x^{d-1}_{\scriptscriptstyle{E}}$ is the Euclidean scalar product. 

There exists a commonly used integral representation of the AdS Schwinger functions called “split representation" \cite{ruhl,pene,pene0,taronna}; it may be obtained by first representing the Legendre function of the second kind $Q$ as a Mehler-Fock transform of the corresponding Legendre function of the first kind $P$ \cite{Bateman,myself} and then expanding $P$ in plane waves on the Lobachevsky manifold \cite{bm,schaeffer}.

The advantage of the representation  (\ref{JJG}) is that it  has admits  a counterpart on the reals that may be  obtained by Wick rotation while the split representation cannot be Wick-rotated (see appendix \ref{aapb}).  Here is the corresponding representation of the Feynman propagator of the Klein-Gordon operator $\Box+m^2_\lambda$ in the real manifold $AdS_M$:
\begin{equation}
G^{(d)}_\lambda(x_1(\x_1,u),x_2(\x_2,u'))= 
\frac{1}{2  } \left(u u'\right)^{\frac{d-1}{2}}
   \int G^{M_{d-1}}_m(\x_1-\x_2) J_{\frac{d-1}{2}+\lambda }\left(m u  \right) J_{\frac{d-1}{2}+\lambda }\left( m  u'\right)dm^2  \label{JJG2}
\end{equation}
where 
\begin{eqnarray}
 && G^{M_{d-1}}_m(\x) =  -\frac{1}{(2\pi)^{d-1 }} \int \frac{e^{-i\k\x}}{{\k^2-m^2+i \epsilon}}  d\k
\end{eqnarray}
is the Feynman propagator of the Klein-Gordon operator $\Box_M+m^2$  in a $(d-1)$-dimensional Minkowski spacetime $M_{d-1}$.
The  verification that  this expression is indeed a propagator is easy but also instructive:
\begin{eqnarray}
&& (\Box+ m^2_\lambda)G^{(d)}_\lambda(x_1,x_2)
  = u^{2} \Box_M G^{(d)}_\lambda(x_1,x_2) - u^d{\partial_u}  ( u^{2-d} \partial_u  G^{(d)}_\lambda(x_1,x_2))+m^2_\lambda G^{(d)}_\lambda(x_1,x_2) =  
\cr  &&\cr  &&  = m^2_\lambda G^{(d)}_\lambda(x_1,x_2)+
 \frac12 u^2 {\left(u u'\right)^{\frac{d-1}{2}}} \delta(\x_1-\x_2) 
  \int_0^\infty   J_{\frac{d-1}{2}+\lambda }\left(m u  \right) J_{\frac{d-1}{2}+\lambda }\left( m  u'\right) dm^2 +\cr && 
- \frac12 m^2 u^2  {\left(u u'\right)^{\frac{d-1}{2}}}
  \int_0^\infty   G^{M_{d-1}}_m(\x_1-\x_2) J_{\frac{d-1}{2}+\lambda }\left(m u  \right) J_{\frac{d-1}{2}+\lambda }\left( m  u'\right) dm^2 +\cr &&  -
\frac12 {\left(u u'\right)^{\frac{d-1}{2}}}
  \int_0^\infty   G^{M_{d-1}}_m(\x_1-\x_2)  \left(\lambda  (\lambda+d -1)-m^2 u^2\right)J_{\frac{d-1}{2}+\lambda }\left(m u  \right) J_{\frac{d-1}{2}+\lambda }\left( m  u'\right) dm^2 \cr&& 
    \cr &&  =
u^d  \delta(\x_1-\x_2) \delta(u,u')
\end{eqnarray}
where we used Eq. (\ref{jlk}) and the inversion theorem of the Hankel transform \cite{batemantransform2}.

The relation of AdS Euclidean diagrams with real space diagrams  \cite{pene1,pene,malda} remains however poorly understood also because calculations based on the split representation cannot be Wick-rotated.  Eqs. (\ref{JJG}) and (\ref{JJG2}) may shed some new light on this issue. 
In particular they immediately implies that banana diagrams  \cite{loopads} and more generally diagrams with no external lines computed in the Euclidean manifold $EAdS_d$ can be Wick-rotated and coincide with the same diagrams calculated in $AdS_M$.

What is less obvious is whether a more general diagram computed by integrating only in a Poincar\'e patch $AdS_M$  produces an AdS invariant result. 
We want to suggest that this might be generally true by computing just  the simplest  one-line diagram  i.e. the integral of the propagator (\ref{JJG2}) over $AdS_M$ w.r.t. one of the two points letting the other point fixed. The calculation in flat space is just trivial:
\begin{eqnarray}
&& 
\int_{M_{d-1}} G^{M_{d-1}}_{m}(\x_1-\x_2)  d\x_1  = -\frac{1}{(2\pi)^{d-1 }} \int \frac{e^{-i\k(\x_1-\x_2)}}{{\k^2-m^2+i \epsilon}} d\x_1  d\k = \frac{1}{m^2}.
\label{diagram1M}
\end{eqnarray}
Using this result we can perform the integral of (\ref{JJG2}) over  $AdS_M$:
\begin{eqnarray}
&& \int_{AdS_M} G^{d}_{\lambda}(x_1, x_2)  \sqrt{g(x_1)}\, dx_1 =
\int_0^\infty \int_0^\infty {\left( \frac{u'}{u}\right)^{\frac{d-1}{2}}}
    \frac{ J_{\frac{d-1}{2}+\lambda }\left(m u  \right) J_{\frac{d-1}{2}+\lambda }\left( m  u'\right)}{m u}    {du} dm= 
\cr && 
= \frac{2^{-\frac{d+1}{2}}  \Gamma
   \left(\frac{\lambda }{2}\right)  }{\Gamma
   \left(\frac{1}{2} (d+\lambda +1)\right)}\int_0^\infty {  {(mu')}^{\frac{d-1}{2}} J_{\frac{d-1}{2}+\lambda }(m u')}\frac{dm}{m} 
\label{diagram}= \frac{1}{\lambda  (\lambda+d -1)}= \frac 1{m^2_\lambda}.\end{eqnarray}
This is, {\em mutatis mutandis}, the same result as in flat space; the independence of the integral on $x_2$ is remarkable and makes the result AdS invariant even though $AdS_M$ is not invariant. 

The corresponding Euclidean calculation, on the contrary, is expected to give an invariant result;  using Eq. (\ref{kgtps}) we get \cite{loopads}   
\begin{eqnarray}
 \int_{EAdS} S^{d}_{\lambda}(z^{\scriptscriptstyle E}_1, z^{\scriptscriptstyle E}_2)  \sqrt{g(z^{\scriptscriptstyle E}_1)}\, dz^{\scriptscriptstyle E}_1 =
  \frac {2 ^{\frac {2 -d}{2}} }{\Gamma \left({\frac {d}{2}}\right)} \int _1^\infty 
{ }{e^{-i\pi\frac {d-2}2}} Q^{\frac {d-2}2}_{\frac {d-2} 2+\lambda}(u) (u^2-1)^{\frac{d-2}4}du = \frac{1}{m_\lambda^2}  \label{0loop}
\end{eqnarray}
i.e. the same result as in the above Lorentzian calculation restricted to $AdS_M$ (see \cite{loopads} for details).
The study of more complicated examples and in particular of Witten diagrams will be presented in a forthcoming paper.
\section{Concluding remarks}
 AdS quantum field theory admits a formulation that is structurally rich as its Minkowskian Wightman counterpart provided one  fully exploits  it on the intrinsic complex geometry of AdS rather than working in specific coordinates. By introducing holomorphic AdS plane waves and identifying suitable relative homology cycles supported on the complex null cone, the scalar two-point function is expressed in a manifestly covariant, coordinate-free integral representation that implements covariance, locality, maximal analyticity in a unified way, while addressing some of the conceptual difficulties associated with closed timelike curves. Its diagonalisation in Poincaré coordinates, encoded in a Källén–Lehmann-type superposition over lower-dimensional Minkowski two-point functions with Bessel kernels, not only yields explicit Schwinger and Feynman propagators but also helps to clarify the status of the Poincar\'e patch and the precise sense in which Euclidean Feynman diagrams on Lobachevski space can be Wick-rotated to diagrams supported entirely in a single AdS patch without loss of AdS covariance. 

Altogether, the framework provides a robust and versatile starting point for perturbative AdS QFT and offers a new  setting in which to revisit and refine holographic constructions that rely on AdS correlation functions and their boundary limits.

\appendix
\begin{appendix}

\section{Algebraic characterization of the chiral cones} \label{Appa}
In order to characterize the chiral cones $\CC_{\pm}$ by algebraic inequalities we  write  $ M = g M_{0d}g^{-1}$;  given a vector $\zeta$ belonging to, say, $\CC_+$ there is a $\xi\in \CC_d$ such that 
\begin{equation} 
 \zeta = 
 e^ {(s+it) M}\,  \xi= g  \ e^{(s+i t)  M_{0d}}  g^{-1} \xi   = 
g  \ e^{(s+\alpha + i t)  M_{0d}}  \xi' =  g' \zeta',
 \label{eq48}
 \end{equation} 
 where  
\begin{equation}
\alpha = \arctan{\frac{(g^{-1} \xi)^0}{(g^{-1} \xi)^d}},\ \  \ \ \xi' =  \left( \begin{array}{c} 0 \cr\vv{ g^{-1} \xi}  \cr \left|\vv{ g^{-1} \xi}  \right| \end{array}\right) \in \CC_d ,\ \ 
\  \ 
 \zeta' = e^{it M_{0d}}  \xi'   \in  \CC_{+} \ .  \label{eq48b}
 \end{equation} 
Inspection shows  that 
\begin{equation}
\scalar{\Im \zeta'}{\Im \zeta'} = |\vv \xi' |^2 \sinh  ^2  t  >0  \ \  \makebox{ and  } \  \ \epsilon(\zeta')= \frac 1 2    |\vv \xi' | \sinh 2t  >0.
\end{equation} 
 By $G_0$-invariance Eq. (\ref{eq48})   implies that also $\scalar{\Im \zeta}{\Im \zeta} >0$. 
Furthermore,  
 for any null complex vector $\zeta=\xi+i \chi \in \CC_d^{(c)}$ there holds the identity 
\begin{eqnarray}
&& \epsilon(\zeta)^2= (\chi^0 \xi^d - \chi^d \xi^0)^2 =(\scalar{\chi}{\chi})(|\vec{\xi}|^2+ |\vec{\chi}|^2) +(\scalar{\chi}{\chi})^2 + |\vec{\xi}|^2 |\vec{\chi}|^2 - (\vec{\xi}\cdot \vec{\chi})^2.
\label{2.5.uu}
\end{eqnarray}
Eq. (\ref{2.5.uu})  shows that $\epsilon(\zeta)^2$ is strictly positive when $\scalar{\chi}{\chi}$ is strictly positive;  since  $G_0$ is connected, Eq. (\ref{eq48}) in turn implies that there is a continuous path connecting $\zeta'$ and $\zeta$ and therefore  also $\epsilon(\zeta)>0$.
Summarizing,  the open cone  
 \begin{eqnarray} \label{chiralcone}
 \CC = \{\zeta=\xi+i\chi \in \CC_d^{(c)}:\
\scalar{\xi}{\xi} =\scalar{\chi}{\chi}>0, \ \scalar{\xi}{\chi} = 0 \}
\end{eqnarray}
is the disjoint union of two $G_0$-invariant subsets
$
 \CC'_{\pm} = \{\zeta=\xi+i\chi \in \CC,\   \epsilon(\zeta)\gtrless 0\}
$ and $\CC_\pm\subset \CC'_\pm$. 
\vskip 10pt
To show the opposite inclusion, let us suppose that $\zeta=\xi+i\chi \in \CC'_+$;  there is a $t>0$ such that  $\scalar{\xi}{\xi}  = \scalar{\chi}{\chi}  = \sinh^2 t$.  
A transformation $g_0\in G_0$ can be found such
that $\xi = g_0\,\sinh(t)\,e_d$ and $\chi = g_0\,\sinh(t)\,e_0$. By posing 
\begin{equation}
g(t) =\left(
\begin{array}{cccc}
 1 & 0 & 0 & 0 \\
 0 & 1_{d-2} & 0 & 0 \\
 0 & 0 & \coth (t) & 1/\sinh t \\
 0 & 0 &1/\sinh t  & \coth (t) \\
\end{array}
\right)
\end{equation}
(note that $g(t) g(t')  \not = g(t+t')$)  we get that 
\begin{eqnarray}
g(t)( i\,  \sinh(t)\,e_0 +e_{d-1}+ \cosh(t)\,e_d )=e^{itM_{0d}}(e_{d-1}+e_d) \label{oip}
\end{eqnarray}
and therefore
\begin{equation}
\zeta= \xi+i\chi  =g_0  g(t)^{-1}  e^{itM_{0d}}(e_{d-1}+e_d) \in \CC_+.
\end{equation}
We may thus identify the chiral cones with the following subsets of $\CC_d^{(c)}$:
\begin{eqnarray} \label{chiralcones+}
 \CC_{\pm} = \{\zeta=\xi+i\chi \in \ambc:\
\scalar{\xi}{\xi}  = \scalar{\chi}{\chi}  >0, \ \scalar{\xi}{\chi}  = 0,  \  \epsilon(\zeta)\gtrless 0\}.
\end{eqnarray}
The above construction also shows that the chiral cones  may be generated simply by acting on the  point $e_{d-1}+e_d$ as follows:
 \begin{eqnarray}
\CC_{\pm} = \{g\, \exp(it
M_{0d})(e_{d-1}+e_d)\ :\
g \in G_0,\ \ \ t \gtrless 0\}.
\label{3.2.1} 
 \end{eqnarray}
By using the complex spherical coordinates of the cone $\CC_d^{(c)}$
\begin{equation}\label{AdSCParam2} \zeta(t+is,\lambda, \vec a+i \vec b)=  \left\{  \begin{array}{l}\lambda \sin (t+i s), \cr \lambda\ (\vec  a+i \vec b), \cr \lambda\cos (t+i s), \end{array} \right. \ \ \ 
|\vec a|^2-|\vec b|^2=1, \ \ \vec a \cdot \vec b = 0
\end{equation}
we see that also $\CC_{-}$ 
and  
$\CC_{+}$ are semi-tubes in ${\bf C}^d$ 
bordered by surfaces $s = s_\pm(|\vec b^2|)$ :
\begin{eqnarray}
&& {\CC_+}\ : \ \ \ \  s > +{\rm arcsh} (|\vec b|)\\
  &&  {\CC_-}\ : \ \ \ \  s <- {\rm arcsh}  (|\vec b|)
\end{eqnarray}
 $\widetilde \CC_{\pm}$  are given by the same formulae  
where the real variable $t$ has been unfolded.

 \label{aapa}
\section{Remarks on the representations of the Euclidean propagator}
\label{aapb}

In section~\ref{kgf}, we showed that certain specific Legendre functions of the second kind, multiplied by $(z^2-1)^{-\frac{d-2}{4}}$, possess analyticity properties ideally suited to the geometry of the anti-de Sitter manifold and its universal covering. In particular, the maximal analyticity of the two-point function~(\ref{kgtp}) in the covering cut-plane $\widetilde{\Delta}_1$ arises from AdS invariance, local commutativity, and the positivity of the energy spectrum, while the location of the cut reflects causal structure.

By contrast, as recalled in section~\ref{kgf}, certain {Legendre functions of the first kind}, multiplied by the same factor,  exhibit analyticity properties naturally adapted to the de Sitter  geometry and are analytic in the cut-plane $\Delta$. 

The above analytic structures may be used to compare the status of the so-called "split representation" of the AdS Euclidean propagator with its Hankel form~(\ref{JJG}). To this end, it is necessary to recall some properties of the plane-wave expansion~\cite{bgm,bm} of the  (Bunch-Davies) de Sitter Wightman function,  keeping the details to a strict minimum.

The complex de Sitter universe may be represented as the manifold
\begin{equation}
dS_d^{(c)} = \{ z \in M_{d+1}^{(c)} : \langle z,z\rangle = -1\},
\label{s.2}
\end{equation}
(we set the radius $R=1$) immersed in a $(d+1)$-dimensional complex Minkowski spacetime. The scalar product\footnote{We use the notation $\langle z_1,z_2\rangle$ for the dS scalar product. This is distinct from the AdS scalar product~(\ref{adsscp}):
\[
z_1\cdot z_2 = z_1^0 z_2^0 - z_1^1 z_2^1 - \dots - z_1^{d-1} z_2^{d-1} + z_1^d z_2^d.
\]
The real manifolds $dS_d$ and $AdS_d$ are real submanifolds of the same complex sphere.} of two events is understood in the ambient Minkowski spacetime sense:
\begin{equation}
\zeta = \langle z_1,z_2\rangle = z_1^0 z_2^0 - z_1^1 z_2^1 - \dots - z_1^d z_2^d.
\label{dssc}
\end{equation}
As complex manifolds, $dS^{(c)}_d$ and $\AdSC$ coincide and are equivalent to the complex $d$-dimensional sphere $S_d^{(c)}$:
\[
dS_d^{(c)} \simeq \AdSC \simeq S_d^{(c)}.
\]
The real de Sitter universe is identified with the one-sheeted hyperboloid
\begin{equation}
dS_d = \{ x \in M_{d+1} : \langle x,x\rangle = -1 \}.
\label{s.2y}
\end{equation}
Two real events $x_1$ and $x_2$ are timelike separated if and only if 
\[
(x_1-x_2)^2 = -2 - 2\, x_1 \cdot x_2 > 0.
\]
The forward lightcone
\begin{equation}
C_+ = \partial V_+ = \{\xi \in M_{d+1} : \xi^2 = \langle \xi,\xi\rangle = 0, \ \xi^0>0\},
\end{equation}
which describes the boundary at timelike infinity of the real de Sitter manifold, may be interpreted as the space of momentum directions of free particles~\cite{cacc}. It is called ``the absolute'' in Gel'fand's classic book~\cite{gelfand}, and has been curiously renamed the ``celestial sphere'' in more recent literature.

The complex de Sitter space-time contains de Sitter-invariant  \emph{tubular domains of analyticity}~\cite{bm}, 
whose topology is much simpler than that of the AdS chiral tubes introduced in section~\ref{general}. 
They can be described simply as the intersections of $dS_d^{(c)}$ with the tubes $T_\pm$ of the ambient complex Minkowski space $M_{d+1}^{(c)}$:
\begin{equation}
{\cal T}_\pm = dS_d^{(c)} \cap T_\pm = \{x+iy \in dS_d^{(c)} : y \in \pm V_+\}.
\label{s.2.1}
\end{equation}
The tubes $\mathcal{T}_-$ and $\mathcal{T}_+$ are bounded by two Riemannian (rather than Euclidean) manifolds, $\mathcal{H}_-$ and $\mathcal{H}_+$, whose points are purely imaginary:
\begin{equation}
\mathcal{H}_\pm=\{x+iy\in dS_d^{(c)}: x=0,\; y^2=\langle y,y\rangle=1,\; \mathrm{sign}\,y^0=\pm\}.
\end{equation}
No analogous structure appears in the Minkowski or AdS tubes, which lack such boundaries away from the real boundary. Both $\mathcal{H}_-$ and $\mathcal{H}_+$ are equivalent to the $d$-dimensional Lobachevsky space $H_d$, i.e.\ the Euclidean AdS manifold.

The simple topology of the tubes is reflected in an equally simple definition of global plane waves, in contrast with the complexity of the corresponding AdS holomorphic waves described in the present paper. Given any complex number $\lambda$,   plane waves in the dS universe are constructed as follows~\cite{bgm,bm}:
\begin{equation}
{\cal T}_\pm \times C_+ \ni (z,\xi) \mapsto \psi^\pm_\lambda(z,\xi) = \langle\xi \, , z\rangle^\lambda.
\label{dswaves}
\end{equation}
They are univalued and holomorphic in the tubular domains ${\cal T}_\pm$ and solve there the Klein-Gordon equation
\begin{equation}
(\Box_{dS} + \mu_\lambda^2)\, \psi^\pm_\lambda = 0
\label{KGnu}
\end{equation} 
with complex squared mass
\begin{equation}
\mu_\lambda^2 = \lambda(1-d-\lambda) = \frac{(d-1)^2}{4} + \kappa^2, \qquad \lambda = -\frac{d-1}{2} + i\kappa.
\label{cmass}
\end{equation}

A plane-wave expansion of the dS two-point function~in terms of the above plane waves has been available for a long time~\cite{bgm,bm}. For any pair of events $z_1, z_2$  in the domain ${\cal T}_- \times {\cal T}_+$, the canonically normalized (so-called) Bunch-Davies de Sitter two-point function is a holomorphic superposition of plane waves, precisely as in (\ref{pwduintnu}) and (\ref{tpm}):
\begin{align}
W^{(dS)}_\kappa(z_1,z_2) &= 
\frac{\Gamma\Big(\frac{d-1}{2}+i\kappa\Big)\Gamma\Big(\frac{d-1}{2}-i\kappa\Big) e^{\pi \kappa}}{2^{d+1} \pi^d} 
\int_{\gamma} \langle\xi \, , z_1\rangle^{-\frac{d-1}{2}-i\kappa} \langle\xi \, , z_2\rangle^{-\frac{d-1}{2}+i\kappa} \, d\mu(\xi),
\label{uup}
\end{align}
where $d\mu_\gamma(\xi)$ is the restriction of the volume form of the cone $C_+$ to a complete $(d-1)$-dimensional cycle\footnote{For example one may chose the spherical section of the cone  canonically oriented:
\begin{equation}
 \gamma_0= C_+ \cap \{\xi\ :\ \xi^0 =1\}
=  \{ \xi \in C_+ : {\xi^1}^2 + \ldots + {\xi^d}^2 =
1\}.
\end{equation}
With this choice $d\mu(\xi)$  coincides
with the rotation invariant measure   normalized 
usual:
\begin{equation}
\omega_{d}=\int_{\gamma_0}d\mu(\xi)  = \frac{2\pi^\frac d2}{\Gamma\left(\frac d2\right)}.
\label{norms}
\end{equation}}
$\gamma$. The homogeneity degree $(1-d)$ of the integrand implies that it is a closed differential form and therefore the result  depends only on the invariant scalar product $\zeta$. A straightforward calculation~\cite{bgm,bm} shows that
\begin{align}
W^{(dS)}_\kappa(z_1,z_2) &= 
\frac{\Gamma\Big(\frac{d-1}{2}+i\kappa\Big)\Gamma\Big(\frac{d-1}{2}-i\kappa\Big)}{2(2\pi)^{d/2}}
(\zeta-1)^{-\frac{d-2}{4}} (\zeta+1)^{-\frac{d-2}{4}} 
P_{-\frac12+i\kappa}^{-\frac{d-2}{2}}(\zeta).
\label{pppds}
\end{align}

As opposed to the AdS case, the two-point function~\eqref{pppds} is invariant under the involution $\kappa \to -\kappa$. It explicitly provides the analytic continuation of~\eqref{uup} to the cut plane $\Delta$, which is the projection of all pairs of complex events of $dS_d^{(c)}$, except for pairs of real events which are timelike separated, whose image forms the causal cut. In particular, $\Delta$ contains the images of all pairs of non-coincident de Sitter Euclidean points.

Eq.~(\ref{pop}) --- which is nothing but an identity among Legendre functions --- may now be rewritten in a more suggestive way in the present context:
\begin{equation}
W^{(d)}_{-\nu} (z_1\cdot z_2) - W^{(d)}_{\nu} (z_1\cdot z_2)
= \sin(\pi\nu)\, W^{(dS)}_{i\nu} (z_1\cdot z_2),
\label{pop2}
\end{equation}
valid for pairs of events $z_1,z_2 \in \AdSC$ such that their AdS scalar product $z_1\cdot z_2$ belongs to the cut plane $\Delta_2$, here regarded as the fundamental sheet $\widetilde{\Delta}_1$.

\subsection{Constructing the split representation of the Euclidean propagator}

We now present the construction of the "split representation" of the AdS Euclidean propagator~(\ref{kgtps}). 
This is done in two steps. In the first step we expand the propagator~(\ref{kgtps}), which is essentially a Legendre function of the second kind, as a superposition of Legendre functions of the first kind. 

This idea was first introduced as a useful technical tool in~\cite{dusedau}. In the two-dimensional case such an expansion has been known for a very long time~\cite[Eq.~(7.213)]{grad}: 
\begin{equation}
Q_{-\frac{1}{2} + \nu}(z) =  \int_0^\infty \kk \, \tanh (\pi \kk) 
 P^{}_{-\frac{1}{2} + i \kk} (z)\,
(\kk^2+\nu^2)^{-1} \ d\kk , \ \ \ \Re \nu >0, \ \ z \in \Delta_2\; \label{grad}.
\end{equation}
The above identity holds, in particular, for real \(z>1\). However, it is clear that a problem arises on the cut \( [-1,1] \). Indeed, the integrand is holomorphic in the cut-plane \(\Delta\), whereas the Legendre function on the left-hand side is defined only for \(z \in \Delta_2 \subset \Delta\). Since the branch cut of \(Q\) extends further, the integral representation becomes ill-defined when \(z\) lies on the cut.

To find the extension of the above formula to general dimension $d$ we may  take the Riemann-Liouville transform of both sides of  (\ref{grad}) \cite{loopads}; the sought extension immediately follows:
\begin{align}
  e^{-i \pi  \frac{d-2}2 }  Q_{-\frac{1}{2}+\nu }^{\frac{d-2}2
   }(z) & = 
   \int_0^\infty\frac{ \kappa\, \sinh (\pi \kappa)  \Gamma \left(\frac{d-1}{2} -i
   \kappa \right) \Gamma
   \left(\frac{d-1}{2} +i \kappa
   \right)}{\pi (\kappa^2+\nu^2)}{} P^{-\frac{d-2}2}_{-\frac{1}{2} + i \kappa} (z)d\kappa. \label{a11}
\end{align}
In our context, by using eqs. (\ref{kgtpnu}) and (\ref{pppds}), the above identity may be rewritten as follows
\begin{align}
W^{(d)}_{\nu}(z_1\cdot z_2) =
\frac 2 \pi \int_0^\infty\frac{  \kk \, \sinh (\pi \kappa)  } {(\kk^2+\nu^2)}W^{(dS)}_\kappa(z_1\cdot z_2)d\kappa \label{weird0}, \ \    z_1,z_2 \in \AdSC, \  z_1\cdot z_2\in 
\Delta_2. 
\end{align}
However this is not the only possible choice to deploy eq. (\ref{a11}). 
Given any two complex AdS events  $z_1$ and $z_2$ such that  $z_1\cdot z_2 \in \Delta_2$, there exist infinitely many pairs of complex  de Sitter events $\tilde z_1$ and $\tilde z_2$ such that $\tilde z_1 \in {\cal T}_-$\, , \  $ \tilde z_2 \in {\cal T}_+$ and
\begin{equation}
 (z_1\cdot z_2)_{AdS} =  \langle \tilde z_1 , \tilde z_2\rangle_{dS}.
\end{equation}
With all the above specifications the following identity holds:
\begin{align}
W^{(d)}_{\nu}(z_1\cdot z_2) =
\int_0^\infty d\kk \frac{e^{\pi \kk}  \Gamma\Big(\frac{d-1}{2}+i\kk\Big)\Gamma\Big(\frac{d-1}{2}-i\kk\Big)  } {(2\pi)^{d}(\kk^2+\nu^2)\Gamma (i \kk ) \Gamma (-i \kk )}
\int_{\gamma} \langle\xi \, , \tilde z_1\rangle^{-\frac{d-1}{2}-i\kk} \langle\xi \, , \tilde z_2\rangle^{-\frac{d-1}{2}+i\kappa} \, d\mu(\xi).
\end{align}
Suppose in particular that  the AdS events are Euclidean: $z_1,z_2 \in EAdS_d$. We may introduce two possible {\em global} maps from the Euclidean AdS manifold to the each of the inner boundaries of the de Sitter tubes as follows: 
\begin{equation}
EAdS_d \ni\left\{\begin{array}{l} y_1^{\scriptscriptstyle{E}}= (iy_1^0, \vec y_1, y_1^d) \to \tilde{y_1}=(-i y_1^d,-i \vec y_1, -i y_1^0) \in {\cal H}_-\ \  (y^d_1>0),\cr \cr 
y_2^{\scriptscriptstyle{E}}= (iy_2^0, \vec y_2, y_2^d) \to \tilde{y_2}=(i y_2^d,i \vec y_2, i y_2^0) \in {\cal H}_+ \ \ \ \ \ \ \ \ \ (y^d_2>0)\end{array}\right.
\end{equation} 
It follows that 
\begin{equation}
z_1^{\scriptscriptstyle{E}}\cdot z_1^{\scriptscriptstyle{E}} =  y_1^d y_2^d -\vec y_1 \cdot \vec y_2 - y_1^0 y_2^0  = \langle \tilde{z_1},\tilde{z_2}\rangle = \cosh v.  \label{splitmap}
\end{equation}

Inserting (\ref{splitmap}) into  Eq. (\ref{weird0}) we finally arrive at the "split representation" \cite{ruhl,pene,pene0,taronna} of the AdS Schwinger propagator:
\begin{eqnarray}
 S^{(d)}_{\lambda}(y_1^{\scriptscriptstyle{E}},y_2^{\scriptscriptstyle{E}}) =  W^{(d)}_{\nu(\lambda)}(y_1^{\scriptscriptstyle{E}}\cdot y_2^{\scriptscriptstyle{E}}) =\frac
{e^{-i\pi\frac {d-2}2}}{(2\pi)^{\frac{d}2}} (\sinh(v))^{-\frac
{d-2}2} Q^{\frac {d-2}2}_{-\frac {1} 2+\nu}(\cosh v) = \cr = 
\int_0^\infty d\kk \frac{e^{\pi \kk}  \Gamma\Big(\frac{d-1}{2}+i\kk\Big)\Gamma\Big(\frac{d-1}{2}-i\kk\Big)  } {(2\pi)^{d}(\kk^2+\nu^2)\Gamma (i \kk ) \Gamma (-i \kk )}
\int_{\gamma} \langle\xi \, , \tilde y_1\rangle^{-\frac{d-1}{2}-i\kk} \langle\xi \, , \tilde y_2\rangle^{-\frac{d-1}{2}+i\kappa} \, d\mu(\xi) \label{kgtpsapp}
\end{eqnarray}
which  has proven to be useful in computing  diagrams on the Euclidean anti de Sitter space. However, it appears from the above construction that a straightforward Wick rotation   to the real anti de Sitter manifold is out of the question.

\end{appendix}

\vskip 10 pt

\section*{Acknowledgments}

I would like to express my gratitude to {Sergio Cacciatori} for his numerous suggestions. I also thank {Charlotte Sleight} and {Massimo Taronna} for several insightful discussions. I thank an anonymous referee  for his  observations, which have helped to improve the manuscript. My deep appreciation goes to the {Institut des Hautes Études Scientifiques (Bures-sur-Yvette)} and the {Physics Department of Fudan University} for their warm hospitality and generous support during the preparation of this paper.

\end{document}